\documentclass[10pt,a4paper]{article}
\usepackage[left=2.5cm, right=2.5cm]{geometry}
\usepackage[utf8]{inputenc}
\usepackage[T1]{fontenc}
\usepackage[spanish]{babel}
\usepackage{comment}
\usepackage{amsmath}
\usepackage{amsfonts}
\usepackage{amssymb}
\usepackage{dirtytalk}
\usepackage{bbm}
\usepackage[]{apacite}
\usepackage{url}
\usepackage{times}
\usepackage{subcaption}
\usepackage{graphicx}
\usepackage{float}
\graphicspath{ {images/} }

\title{Valores extremos de la tasa de inflacion en Costa Rica}
\author{Daniel Aguilar and Breyner Chacón}
\date{7 de diciembre, 2023}

\begin{document}

\bibliographystyle{apacite}

\maketitle

\section{Introducci\'on}
Mantener niveles de inflaci\'on bajos, no negativos y estables es una condici\'on necesaria para la estabilidad de la econom\'ia en su conjunto, por que las autoridades monetarias de la mayor aprte de pa\'ises industrializados, incluyendo el Banco Central de Costa Rica desde el 2005, han orientado su pol\'itica monetaria justo a ese cometido. A\'un as\'i, tanto en Costa Rica como internacionalmente la mayor parte de la modelaci\'on estad\'istica de la inflaci\'on se ha limitado a modelar su esperanza condicionada a distintas covariables bajo el uso de modelos lineales. Ello implica un desconocimiento de la din\'amica de los \emph{valores extremos} de la tasa de inflaci\'on y c\'omo estos se relacionan con otras variables macroecon\'omicas. En Costa Rica esto es de particular importancia ya que en varios periodos recientes se han experimentado tasas de inflaci\'on intertrimestral negativas, lo cual puede ser problem\'atico si ello se torna en un fen\'omeno recurrente. Por ello, en este trabajo se plantea responder cu\'al es la relaci\'on entre la brecha del PIB, expectativas de inflaci\'on, tasa de inflaci\'on importada, y los valores extremos de la tasa de inflaci\'on en Costa Rica. Es decir, se plantea como objetivo principal determinar la relaci\'on entre los valores extremos de la de la tasa de inflaci\'on, brecha del PIB, expectativas de inflaci\'on e inflaci\'on importada.

\section{Marco te\'orico}
\subsection{La inflaci\'on en la teor\'ia econ\'omica moderna}
En la actualidad, toda discusi\'on sobre los determinantes y relaciones de la tasa de inflaci\'on se centra en un grupo de modelos crucial: las curvas de Phillips. En su versi\'on actual, la tasa de inflaci\'on depende postivamente de dos factores fundamentales: alguna medida de la actividad econ\'omica real y las expectativas de inflaci\'on. La conjunci\'on de estos dos factores como determinantes del fen\'omeno inflacionario se puede comprender como el resultado de un proceso hist\'orico en la evoluci\'on de la teor\'ia econ\'omica. Por ello en esta secci\'on se expone primero una breve s\'intesis de la evoluci\'on de las teor\'ias de inflaci\'on para luego exponer la teor\'ia vigente. 

\subsubsection{La curva de Phillips}
El primer modelo inflacionario adoptado ampliamente por la comunidad acad\'emica es el modelo presentado por \cite{Phillips}. En este la inflaci\'on $\pi_t$ depende negativamente de la tasa de desempleo $u_t$ de la forma:
\begin{equation}\label{pc}
    \pi_t=\lambda u_t^{-\alpha} +\beta_0; \ \lambda, \alpha>0
\end{equation}
Esta relaci\'on se atribuy\'o a la idea de que, a mayor exceso de oferta en el mercado laboral, i.e., a mayor desempleo, mayor deb\'ia ser el valor absoluto del cambio porcentual en los salarios y, en la medida en la que estos influencian los precios, menor deb\'ia ser la tasa de inflaci\'on. El caracter no lineal de (\ref{pc}) se atribu\'ia a la rigidez a la baja de los salarios nominales\footnote{Este era un fen\'omeno cuya recurrencia era aceptada.}: aumentos en la tasa de desempleo ten\'ian un menor efecto sobre los salarios en la medida en la que estos se \say{resist\'ian} a disminuir. Durante la d\'ecada de 1960 esta fue la teor\'ia predominante sobre inflaci\'on y \ref{pc} disfrut\'o de cierto \'exito emp\'irico.

\subsubsection{Expectativas y tasa de desempleo natural}
A finales de los 60's, la curva de Phillips sufri\'o una importante cr\'itica te\'orica planteda por \cite{Friedman}. La curva de Phillips de entonces implicaba la existencia de un \say{men\'u} de combinaciones entre inflaci\'on y desempleo a disposici\'on de los bancos centrales. Por ello, se argumentaba que la autoridad monetaria pod\'ia disminuir el desempleo v\'ia pol\'itica monetaria: aumentar la masa de dinero generar\'ia inflaci\'on y con ello, como implica (\ref{pc}), el desempleo disminuir\'ia \cite{gordon11}. Para \cite{Friedman} este no era el caso por cuanto la relaci\'on en (\ref{pc}) era inestable ya que omt\'ia el rol de expectativas de los agentes. Estos, al enfrentarse a altas tasas de inflaci\'on, cambiar\'ian sus expectativas y as\'i tambi\'en sus conductas de forma tal que la tasa de inflaci\'on efectiva se presionase al alza\footnote{Por ejemplo, aumentos en la expectativas de inflaci\'on har\'ian que los trabajadores exigiesen salarios m\'as altos, subiendo as\'i los costos de las empresas y con ello los precios. Adicionalmente, cualquier aumento en la demanda generado por la disminuci\'on de los tipos de inter\'es que acompa\~{n}an a una expansi\'on de la oferta monetaria ser\'ia ef\'imero al ver los agentes un alza en la inflaci\'on}. Esto implicaba que para todo momento $t$ exist\'ia una tasa de desempleo \say{natural} $u_t^N$ cuyo valor era \emph{independiente} de la inflaci\'on y oferta monetaria. \\
\ La inestabilidad que esta crítica atribuía (\ref{pc}) se vio emp\'iricamente validada en la d\'ecada de los 70's cuando, en un contexto de importantes aumentos en los precios del petr\'oleo, se dieron importantes alzas en la inflaci\'on en occidente acompa\~{n}adas de decrecimiento en la producci\'on y alto desempleo. Todo esto motiv\'o a que se extendiera la curva de Phillips a una versi\'on que puede caracterizarse de la siguiente forma:
\begin{equation} \label{PC_exp}
    \pi_t = \beta E_{t-1}\{\pi_t\} + \lambda (u_t-u_t^N) + \gamma \cdot z_t; \ \ \gamma, \lambda, \beta. 
\end{equation}
con $E_{t-1}\{\pi_t\}$ la inflaci\'on esperada en el periodo previo y $z_t$ un vector de variables representativas de shocks de oferta, (e.g. precio del petr\'oleo) y $\gamma$  un vector de par\'ametros. Como las expectativas se consideraban dependientes de la inflaci\'on ya observada por los agentes, usualmente se tomaba $E_{t-1}\{\pi_t\}$ como una combinaci\'on lineal de rezagos en $\pi_t$, particularmente en la modelaci\'on econom\'etrica. Adicionalmente se sol\'ian incluir rezagos en la brecha en la tasa de desempleo, la cual se pod\'ia sustituir por la brecha en el PIB \cite{gordon11}. De esta manera se incorporaron los dos factores fundamentales ya mencionados: una medida de la acvtidad econ\'omica real, dada por $u_t-u_t^N$, y las expectativas. Este fue el modelo predominante en la d\'ecada de los 80's y 90's en c\'irculos acad\'emicos y modelaci\'on estad\'istica de los determinantes de la inflaci\'on. 

\subsubsection{Modelos de formaci\'on de precios de la nueva macroeconom\'ia keynesiana}
Si bien el modelo planteado en (\ref{PC_exp} ) present\'o cierto \'exito emp\'irico desde su postulaci\'on hasta la actualidad, carece de un fundamento te\'orico basado en el comportamiento de los agentes individuales que habitan la econom\'ia, lo cual, siguiendo la influyente cr\'itica de \citeA{lucas_crit}, se ha convertido en una exigencia dentro de la econom\'ia ortodoxa. Ello llev\'o al desarrollo de modelos te\'oricos que admit\'ian relaciones similares a (\ref{PC_exp}) deducidas de las decisiones de agentes racionales que toman decisiones \'optimas sujetos a las restricciones impuestas por su entorno econ\'omico. De esto surgieron teor\'ias sobre formaci\'on de precios que, incorporando rigideces al cambio de precios, permiten la deducci\'on de la llamana curva de Phillips de la Nueva Macroeconom\'ia Keynesiana (NKPC), la cual es parte del actual paradigma sobre fluctuaciones econ\'omicas. A fin de ilustrar esto, ac\'a se expone el modelo de formaci\'on de precios de Calvo, que es la opci\'on usual encontrada en la literatura \cite{gali2015monetary}. En este modelo la econom\'ia est\'a poblada por un continuo de firmas indexadas por $i \in [0,1]$ que se diferencian únicamente por el bien que producen y su historial de precios. Cada periodo, solo una fracción $1-\theta$ aleatoria de las firmas puede cambiar su precio. Para cada firma, en cada periodo, la probabilidad de pertenecer a una u otra fracción es independiente de su historial de precios. Aquellas firmas que pueden cambiar el precio fijan un precio que optimiza una sumatoria de los valores presentes esperados de los beneficios futuros. De ello y de la condici\'on de equilibrio en el resto de la econom\'ia se obtiene la NKPC
\begin{equation}\label{NKPC}
    \pi_t=\beta E_t\{\pi_{t+1}\}+\lambda y_t
\end{equation}
donde $E_t\{\pi_{t+1}\}$ son las expectativas en $t$ respecto de $\pi_{t+1}$, $y_t$ es la desviaci\'on logar\'itmica del PIB respecto de su nivel natural\footnote{Es decir, $y_t=ln(Y_t)-ln(Y_t^N)$ con $Y_t^N$ el PIB natural}, $\beta>0$ es un factor de descuento \emph{sobre la utilidad de los hogares} y $\lambda=\theta^{-1}(1-\theta)(1-\beta \theta)>0$. De esta manera la inflaci\'on en $t$ depende (positivamente) de las expectativas que hoy se tienen respecto de la inflaci\'on futura y del exceso en la producci\'on agregada sobre las capacidades de la econom\'ia. 
La NKPC en (\ref{NKPC}) constituye el modelo b\'asico sobre inflaci\'on en la actual ortodoxia econ\'omica y se utiliza como punto de partida para posibles extensiones. Crucialmente, la NKPC contrasta con (\ref{PC_exp}) en el hecho de que son las expectativas \emph{actuales sobre la inflaci\'on presente} y no las expectativas previas sobre la inflaci\'on de hoy \cite{GALI1999195}. 

\subsection{Valores extremos de la inflaci\'on y teor\'ia econ\'omica}
En la secci\'on previa se describi\'o la teor\'ia econ\'omica b\'asica sobre inflaci\'on en la actualidad. En la medida en la que este trabajo trata sobre extremos de tal variable, es necesario considerar c\'omo la teor\'ia econ\'omica trata en particular tal fen\'omeno, para lo cual ac\'a se exponen dos principales teor\'ias.
\par 
Primero, la teor\'ia esbozada previamente puede explicar la presencia de valores extremos en la tasa de inflaci\'on. En particular, ante \emph{shocks} de oferta, como aumentos s\'ubitos del precio de hidrocarburos la teor\'ia anterior predice aumentos en la inflaci\'on; lo mismo para tipos de inter\'es internacionales y condiciones de cr\'edito en el pa\'is. Similarmente, la teor\'ia da un rol importante a las expectativas las cuales muchas veces se postulan como dependientes del entorno econ\'omico y, en particular, de la inflaci\'on observada en el pasado. De ello se sigue que la presencia de valores extremos en la inflaci\'on tienen el potencial de reproducirse. Por otro lado, las relaciones que la teor\'ia b\'asica establece puede usarse para ver c\'omo responden los valores extremos de la distribuci\'on de la inflaci\'on a cambios en el entorno econ\'omico. Por ejemplo,  \cite{lopez_iar} estudian las colas de la distribuci\'on condicional de la tasa de inflaci\'on de Estados Unidos y la eurozona utilizando regresi\'on cuant\'ilica especificada de la forma:
\begin{equation}\label{qr_lopez}
    Q_\tau(\Bar{\pi}_{t,t+4})=(1-\lambda_\tau)\pi_{t-1}^*+\lambda_\tau\pi_t^{LTE}+\theta_\tau(u_t-u_t^N)+\gamma_\tau(\pi_t^I-\pi_t)+\delta_\tau cs_t
\end{equation}
con $\Bar{\pi}_{t,t+4}$ la inflaci\'on intearanual entre los trimestres $t \ \text{y} \ t+4$, $\pi_{t-1}^*$ la inflaci\'on interanual promedio de los cuatro trimestres previos, $\pi_t^{LTE}$ las expectativas de inflaci\'on a largo plazo y $cs_t$ el diferencial de cr\'edito entre bonos corporativos y de gobierno. Este modelo constituye una versi\'on de la NKPC extendida y aplicada sobre los cuantiles. De su trabajo resaltan dos resultados. Primero, para todos los par\'ametros expecto $\gamma$, encuentra una diferencia significativa entre los valores que toman en los cuantiles del $10\% y 50\%$, lo que sugiere que un efecto asim\'etrico de las variables en cuesti\'on sobre el fen\'omeno inflacionario. Segundo, encuentran diferencias importantes en la forma de las distribuciones predictivas estimadas para momentos particulares de su muestra. En particular, para varios trimestres en el 2008 y 2009, tras la crisis financiera de entonces, las densidades presentaban importante probabilidad de deflaci\'on y se encontraban a la izquierda de las correspondientes a trimestres de 1999 y 1995.  
\par
Segundo, se tiene la teor\'ia de deuda-deflaci\'on, que puede resumirse como un c\'irculo de causalidad entre variables econ\'omicas que genera deflaci\'on de forma sistem\'atica. En s\'intesis, la teor\'ia indica que, episodios sostenidos de deflaci\'on generan aumentos de la deuda \emph{real} de los agentes (ya que el monto de la deuda contraida previamente no cambia pero s\'i bajan sus ingresos) lo que puede generar disminuciones en la producci\'on, lo cual, a su ves, presiona los precios a la baja, generando a\'un m\'as deflaci\'on y perpetuando el fen\'omeno. De esta forma la deflaci\'on no solo se puede autopropagar v\'ia expectativas como se vio antes pero tambi\'en v\'ia el valor de la deuda\footnote{En Costa Rica esto es bastante importante dado el nivel de endeudamiento de los hogares.}. Para que se d\'e este proceso retroalimentativo no es necesario que ocurra deflaci\'on \say{de la nada}, lo cual no es usual, sino que una recesi\'on puede constituir el desencadenante del ciclo vicioso.  

\section{Caracter\'isticas generales de los datos y fuentes de informaci\'on}
La poblaci\'on de estudio viene dada por el proceso generador de datos de la tasa de inflaci\'on intertrimestral de Costa Rica. La muestra tomada corresponde a los datos del primer trimestre del 2000 al segundo del 2023 \cite{BCCR2022}, lo cual se hace por el hecho de que previo a tales fechas, no se cuenta con valores de la variable expectativas de inflaci\'on e inflaci\'on de materias primas. Todos los datos crudos fueron obtenidos directamente de la p\'agina web del Banco Central de Costa Rica (BCCR). Para estimar el valor del PIB natural se utiliz\'o la tendencia extraida con el filtro Hodrick-Prescott (HP), como es usual en parte importante de la literatura.

\section{An\'alisis exploratorio de datos}

\subsection{Medidas b\'asicas}


En el cuadro \ref{table: t1} se muestran las medidas b\'asicas de dispersi\'on y tendencia central de las variables. Destaca el hecho de que, en promedio, la brecha en el PIB ha sido negativa, lo cual sugiere una econom\'ia relativamente desacelerada y, por ende, con presiones deflacionarias por parte del sector real de la econom\'ia. La inflación por su parte ha sido positiva al menos en el 75\% de los registros, sin embargo es de notar que existen minimos negativos de esta inflación. Por otro lado en los cuadros \ref{table: t2} y \ref{table: t3} se tienen las estadísticas anterior y posterior a la crisis subprime del 2008 que representó un posible cambio estructural tanto en inflación como en brecha del producto interno bruto, en donde se pueden apreciar que hay cambios en la mediana y media siendo valores mayores en el primer reciente, donde el m\'imino no es negativos, mientras que en el segundo s\'i lo es.  

\begin{table}[H] 
\centering
\begin{tabular}{rlllllll}
  \hline
 & Mínimo & Primer cuartil & Mediana & Media & Tercer cuartil & Máximo \\ 
  \hline 
     Inflación & -0.009246   &  0.004372   & 0.013609   & 0.014649   &  0.023351   & 0.048351   &  \\ 
      Expectativa inflación & 0.000296   & 0.000383   & 0.000621   & 0.000635   & 0.000760   & 0.001274   &    \\ 
    Brecha PIB & -0.102319   & -0.021706   & 0.001631   & -0.001224   & 0.018782   & 0.056714   &  \\ 
    Brecha inflación & -0.600245   & -0.056064   & 0.006304   & -0.004219   & 0.064201   & 0.260430   &  \\
    Inflación Importada & -0.58186   & -0.04122   & 0.01950   &  0.01043   & 0.07521   & 0.25992  &\\
   \hline
\end{tabular}
\caption{Estadísticas descriptivas.}
\label{table: t1}
\end{table}

\begin{figure}[H]
    \centering
    \caption{\\[0.0001cm] \small \textbf{Medidas de dependencia y correlaci\'on entre inflaci\'on intertrimestral y covariables}}
    \includegraphics[scale=0.35]{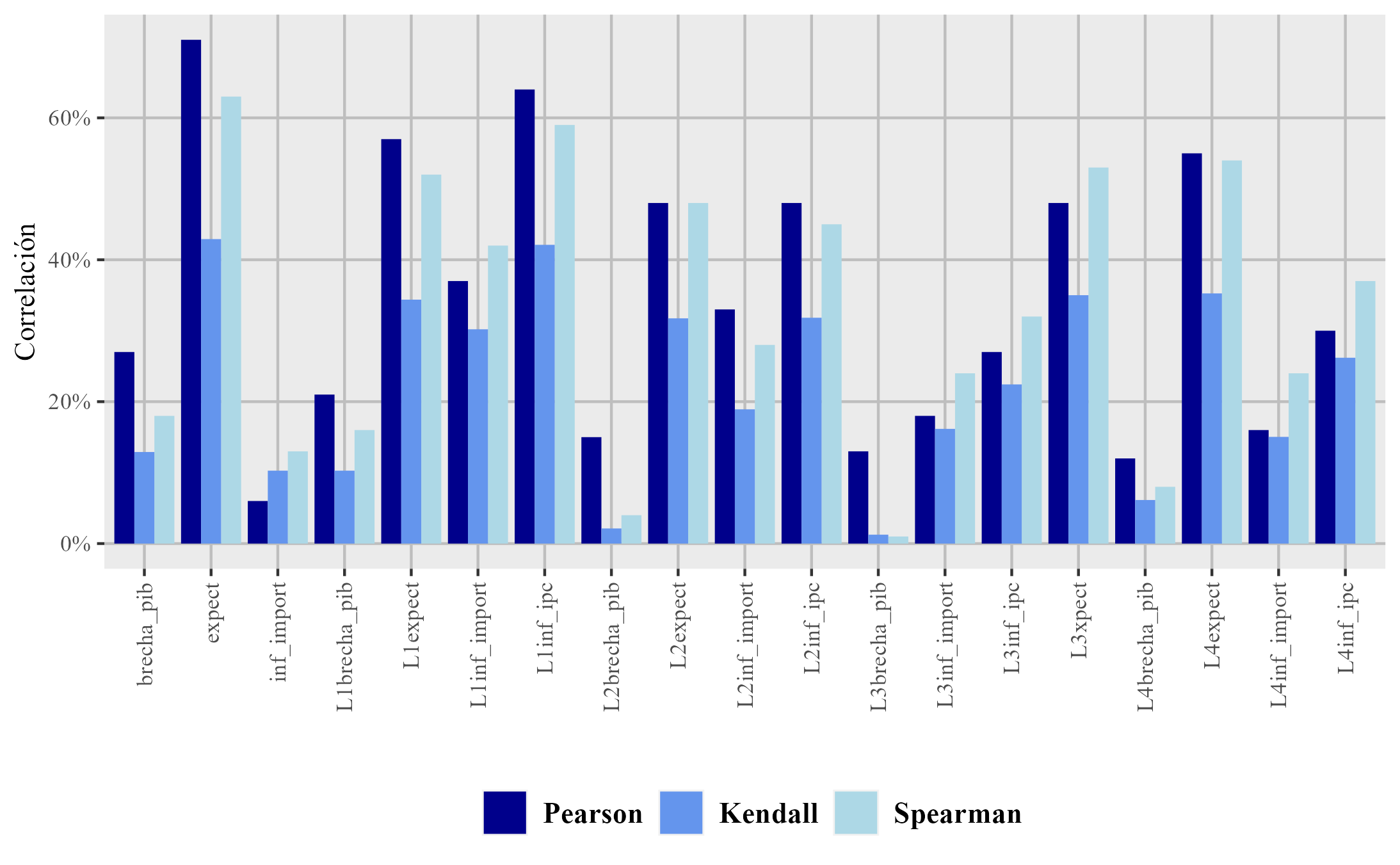}
    \label{fig:corrs_full}
\end{figure}

 \begin{table}[H]
 \centering
 \begin{tabular}{rllllll}
   \hline
  & Mínimo & Primer Cuartil & Mediana & Media & Tercer Cuartil & Máximo \\ 
   \hline
    Inflación & 0.009143   & 0.021250   & 0.024394   & 0.026493   & 0.031357   & 0.048351   \\ 
     Brecha PIB & -0.0477078   & -0.0284269   & 0.0035976   & 0.0005334   & 0.0256752   & 0.0567137   \\ 
    \hline
 \end{tabular}
 \caption{Estadísticas de inflación y brecha del PIB, 2000-2008}
 \label{table: t2}
 \end{table}
 \begin{table}[H]
 \centering
 \begin{tabular}{rllllll}
   \hline
  & Mínimo & Primer Cuartil & Mediana & Media & Tercer Cuartil & Máximo \\ 
   \hline
    Inflación & -0.009246   & 0.001078   & 0.005413   &  0.007503   & 0.011644   & 0.047598   \\ 
     Brecha PIB & -0.1023185   & -0.0169941   & -0.0001721   & -0.0022852   & 0.0171870   & 0.0463569   \\ 
    \hline
 \end{tabular}
 \caption{Estadísticas de inflación y brecha del PIB, 2009-2023}
 \label{table: t3}
 \end{table}
Por otro lado, a fin de explorar la dependencia de la inflaci\'on y las covariables, en la figura \ref{fig:corrs_full} se muestran las correlaciones de Pearson, Kendall y Spearman entre la primera y las segundas, incluyendo el valor contempor\'aneo y hasta cuatro de los rezagos de estas \'ultimas. Para todas los casos, los estad\'isticos mostrados presentan valores positivos, lo que sugiere que estas tienen una influencia directa sobre la inflaci\'on, como indica tambi\'en la teor\'ia. La brecha del PIB presenta mayor dependencia en su valor contempor\'aneo que en los rezagos, siendo el segundo y tercero no inferiores al 20\% bajo los tres estad\'isticos; a\'un as\'i su valor contempor\'aneo y primer rezago parecen tener una dependencia d\'ebil, superando solo el 20\% bajo la medida de correlaci\'on de Pearson. En el caso de las expectatiavs de infalci\'on su dependencia es mucho mayor, siendo m\'as fuerte en el valor contempor\'aneo que en los rezagos, si bien estos parecen tener tambi\'en bastante influencia sobra la inflaci\'on. La inflaci\'on de materias primas importadas presenta medidas de dependencia bajas en su valor contempor\'aneo mientras que, para el primer y segundo rezagos estas se vuelven bastante altas y, para el tercero y cuarto vuelven a niveles bajos. Esto es de esperar por cuanto es razonable asumir que existe un rezago entre la compra de insumos en los mercados internacionales y el subsecuente efecto de este sobre los precios de los bienes de consumo final. Finalmente, la tasa de inflaci\'on presenta alta dependencia con sus autos rezagos, particularmente el primero y segundo. 
\subsection{Evoluci\'on de variables}
\begin{figure}[H]
\centering
\caption{\\[0.0001cm] \small \textbf{Tasa de inflación interimestral y expectativas de inflaci\'on interanuales, 2000-2023}}
    \begin{subfigure}[b]{0.47\textwidth}
        \centering
        \includegraphics[width=\textwidth]{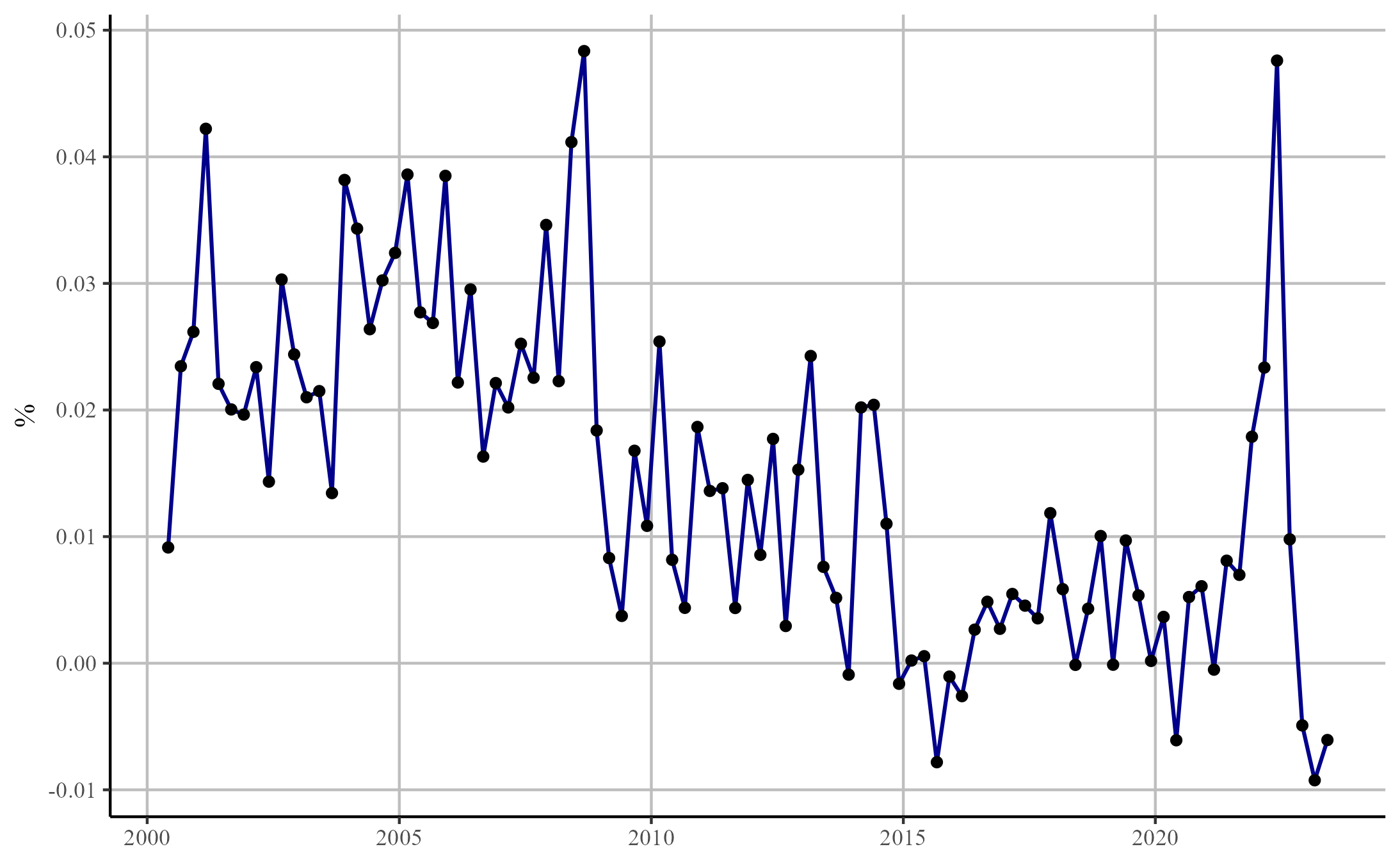}
        \caption{Tasa de inflaci\'on}
        \label{fig:inf}                 
    \end{subfigure}
    \hfill
    \begin{subfigure}[b] {0.47\textwidth}
    \centering
        \centering
        \includegraphics[width=\textwidth]{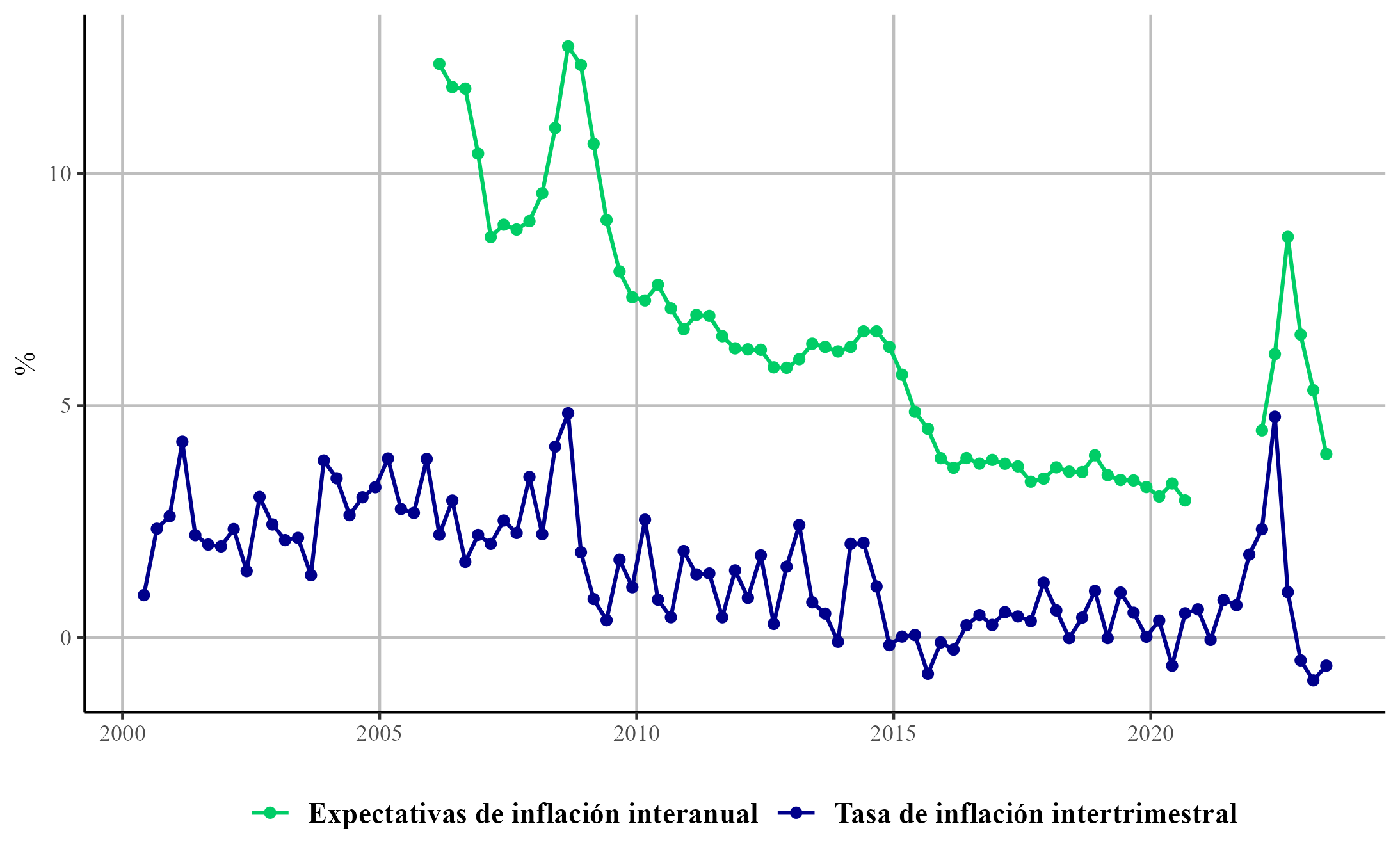}
        \caption{Expectativas de inflaci\'on}
        \label{fig:esperada}
    \end{subfigure}
\end{figure}

En esta subsecci\'on se examina la evoluci\'on de cada una de las covariables en conjunto con la tasa de inflaci\'on. Priemro, en el grafico \ref{fig:inf} se muestra la evoluci\'on de la tasa de inflaci\'on. Destacan dos cosas. Primero, se pueden distinguir dos periodos: previo al 2009 se observan valores m\'as altos y con una mayor volatilidad, mientras que, de ese a\~{n}o en adelante la tasa de inflaci\'on disminuye a valores cercanos al 1\% y su variabilidad baja. Segundo, se presentan varios trimestres con deflaci\'on alrededor del 2015 \footnote{En particular, el cuarto semestre del 2013 y 2014, el tercer y cuarto semestre del 2015 y primero del 2016, el segundo del 2018 y primero del 2019, presentan tasa de inflaci\'on negativa.} y luego del 2020, incluyendo los dos primeros trimestres del 2023.  Similarmente, en la figura \ref{fig:esperada} se observa la tasa de inflaci\'on y las expectativas de inflaci\'on, donde destaca el hecho de que ambas variables presentan disminuci\'on importante luego del 2009 lo que sugiere una respuesta de las expectativas de los agentes ante los cambios en la tasa de inflaci\'on observada. 

\par
\begin{figure}[H]
\centering
\caption{\\[0.0001cm] \small \textbf{Brecha del PIB trimestral e inflaci\'on intertrimestral de materias primas importadas, 2000-2023}}
    \begin{subfigure}[b]{0.47\textwidth}
        \centering
        \includegraphics[width=\textwidth]{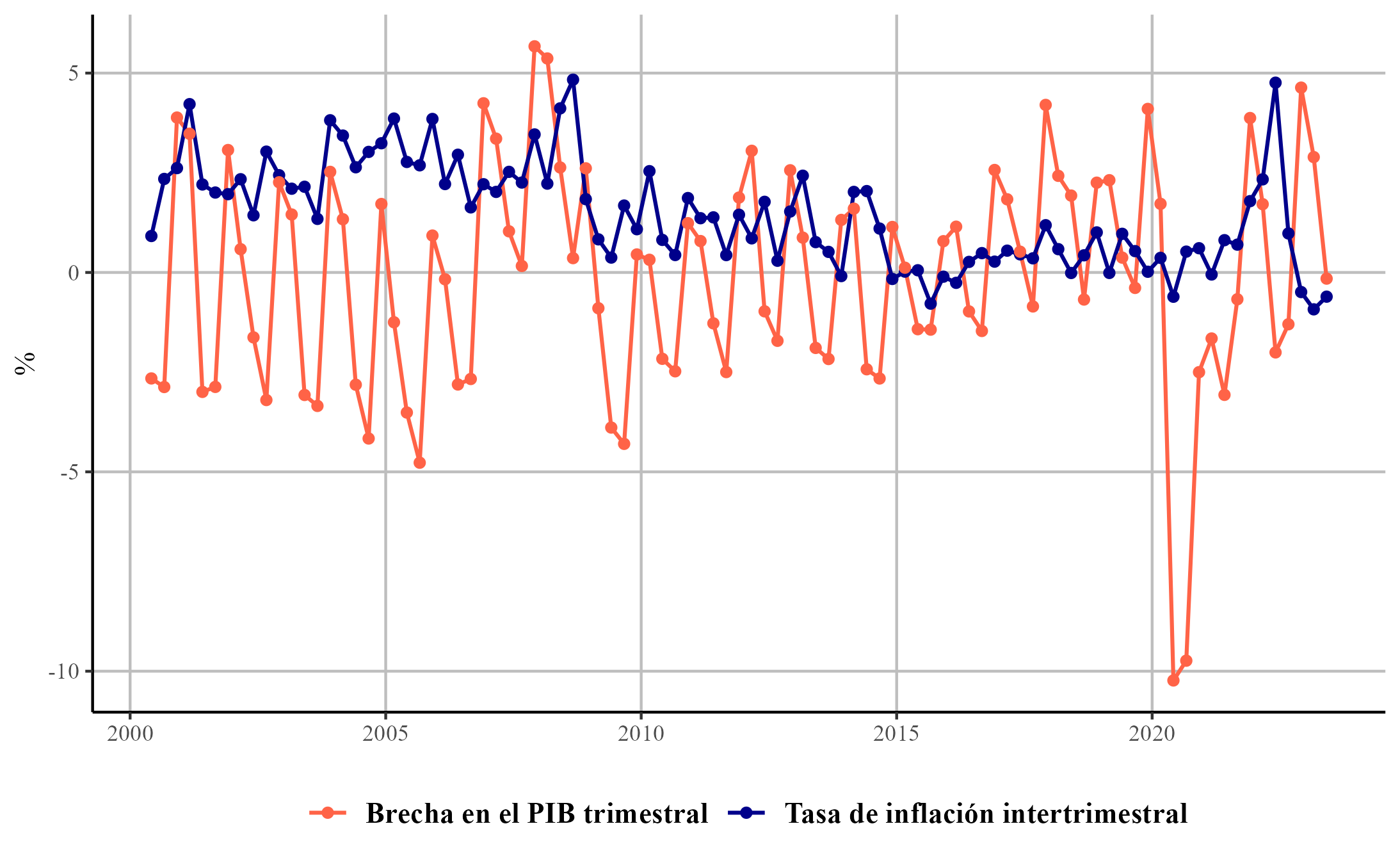}
        \caption{Brecha del PIB}
        \label{fig:brecha}                
    \end{subfigure}
    \hfill
    \begin{subfigure}[b] {0.47\textwidth}
    \centering
        \centering
        \includegraphics[width=\textwidth]{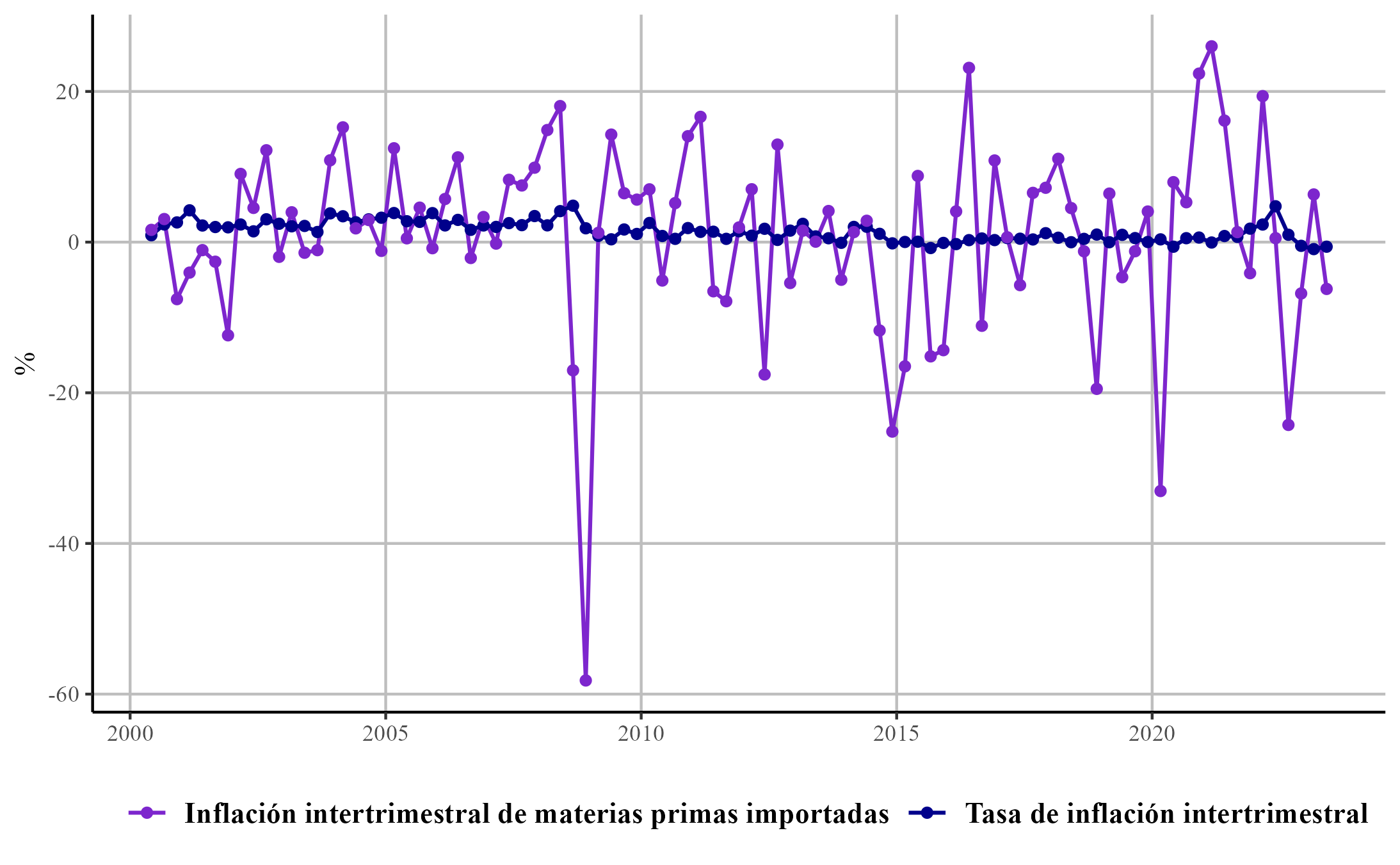}
        \caption{Inflaci\'on importada}
        \label{fig:importada}
    \end{subfigure}
\end{figure}

 En la figura \ref{fig:brecha} se muestra la tasa de inflaci\'on y brecha del producto. Destacan tres observaciones. Primero, la brecha del producto parece tener un importante componente estacional con importantes subidas en el priemr trimestre del a\~{n}o. Segundo, la tasa de inflaic\'on parece tener un comportamiento proc\'icliclo en el sentido de presentar variaciones en la misma direcci\'on que la brecha en el PIB, lo cual coincide con las medidas de correlaci\'on obtenidas. Tercero, exceptuando las observaciones del 2020 y 2021, los trimestres con deflaci\'on han estado acompa\~{n}ados de brechas en el PIB no negativas o negativas pero relativamente cercanas a cero. Esto destaca particularmente si se comparan tales valores de la brecha con los observados en varios trimestres pre 2009, en los que la brecha es negativa y bastante menor, pero no se observa deflaci\'on. Esto sugiere la posibilidad de efectos asim\'etricos de la brecha sobre la tasa de inflaci\'on, siendo mayor la presi\'on que le ejerce al subir que al bajar. 

\par En la figura \ref{fig:importada} se muestra la tasa de inflaci\'on de materias primas importadas y tasa de inflaci\'on dom\'estica. Para todo el periodo las materias primas importadas presentan una mayor volatilidad en el cambio de sus precios y presentan valores absolutos m\'as altos que la inflaci\'on dom\'estica. Se pueden observar fuertes periodos de deflaci\'on en las importaciones de materias primas alrededor del 2015 lo cual podr\'ia explicar la deflaci\'on observada en los precios dom\'esticos en ese mismo periodo. 

\subsection{Extremos en la tasa de inflaci\'on} \label{descrip_extremos}
\begin{figure}[H]
\centering
\caption{\\[0.0001cm] \small \textbf{Medidas de dependencia y correalci\'on entre inflaci\'on intertrimestral y covariables para submuestras}}
    \begin{subfigure}[b]{0.47\textwidth}
        \centering
        \includegraphics[width=\textwidth]{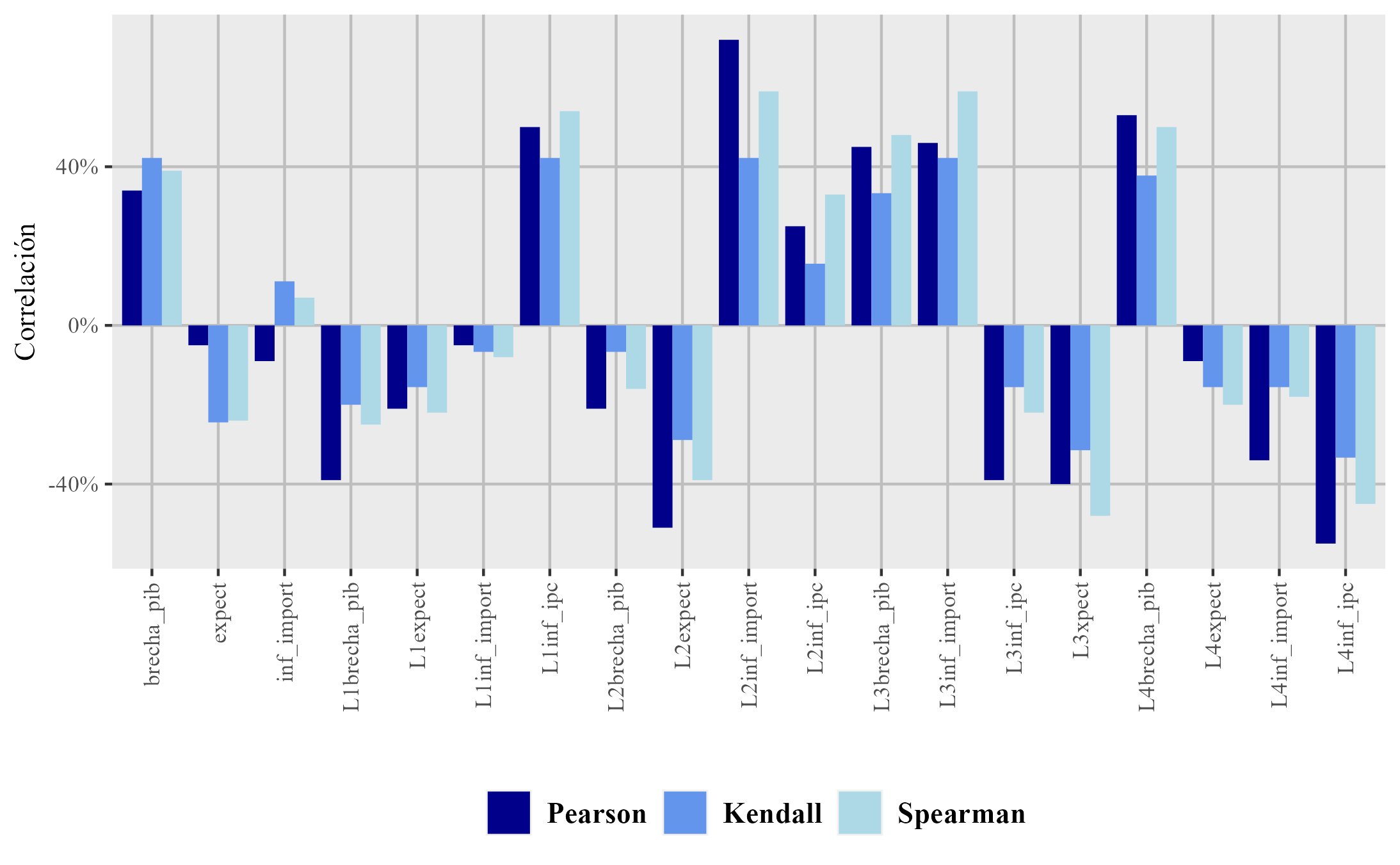}
        \caption{Submuestra con deflaci\'on}
        \label{fig:corr_def}                
    \end{subfigure}
    \hfill
    \begin{subfigure}[b] {0.47\textwidth}
    \centering
        \centering
        \includegraphics[width=\textwidth]{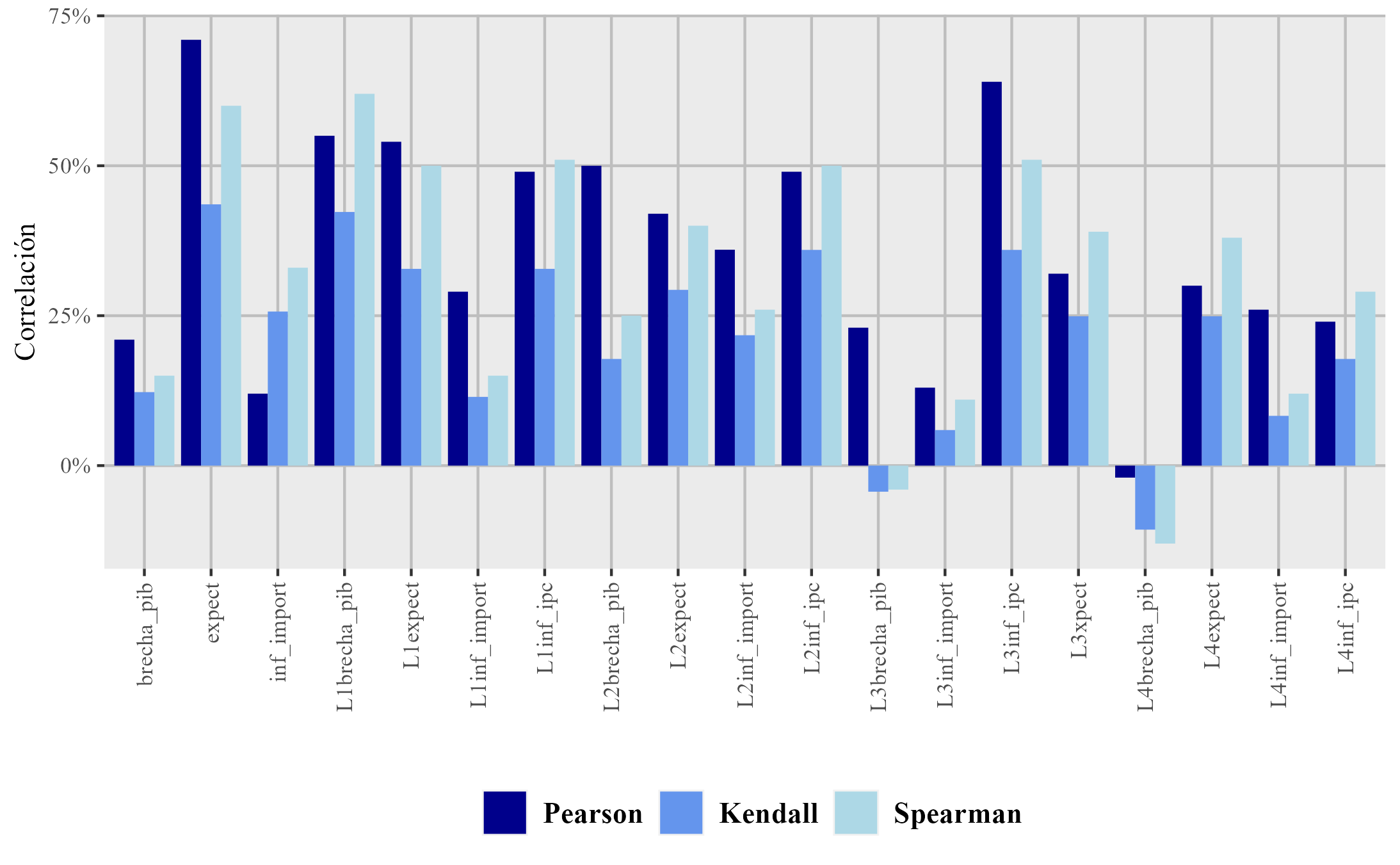}
        \caption{Submuestra con inflaci\'on >1\%}
        \label{fig:corr_hiper}
    \end{subfigure}
    \label{fig:corrs_submuestras}
\end{figure}

En esta subsecci\'on se profundiza en el an\'alisis descriptivo a fin de contemplar las posibles relaciones de las covariables con los extremos de inflaci\'on. En la figura
\ref{fig:corrs_submuestras} se presentan las medidas de correlaci\'on y dependencia para dos submuestras, una con los periodos de deflaci\'on y otros donde la inflaci\'on supera el 1\% \footnote{Las medidas obtenidas son casi id\'enticas para la submuestra con tasa de inflaci\'on superior al 2\%.}. En el segundo caso la medidas son muy similares a las de la muestra completada observadas en la figura \ref{fig:corrs_full}. Sin embargo, en el caso con deflaci\'on existen dos importantes diferencias. Primero, las expectativas de inflaci\'on, en su valor contempor\'aneo y todos sus rezagos, presentan ahora una dependencia negativa y menor al de la muestra completa. Lo mismo sucede con el segundo y tercer rezago en la brecha del PIB y el valor contempor\'aneo y primer rezago de la inflaci\'on importada. Segundo, el tercer y cuarto rezago de la brecha del PIB, as\'i como el segundo y tercero de la inflaci\'on de materias primas importadas, se mantiene positivos y su valor aumenta de forma importante. Todo esto se puede interpretar de la siguiente manera. Estas \'ultimas covariables parecen ser responsables de \emph{shocks} deflacionarios\footnote{En efecto, presentan una dependencia positiva y alta con la inflaci\'on que se manifiesta \'unicamente en periodos con deflaci\'on.}. Por otro lado,  las primeras, parecen influenciar a la tasa de inflaci\'on en general a trav\'es de todo el ciclo econ\'omico y de forma positiva. Ante un shock deflacionario, estas segundas sigue su comportamiento \say{usual} pero la inflaci\'on disminuye en direcci\'on contraria y, por ello, los estad\'isticos de las segundas se recogen negativos y en menor valor absoluto. 

\section{Metodolog\'ia}
En la literatura estad\'istica es com\'un modelar la ocurrencia de extremos utilizando herramientas de la teor\'ia de valores extremos como modelos de excesos sobre umbrales y de m\'aximos de bloque \cite{coles2001introduction}, en los que, para series de tiempo, se asume estacionariedad del proceso que les genera. En el contexto de este trabajo este supuesto es altamente cuestionable, por lo que, siguiendo a \cite{lopez_iar} se ha decidido aplicar regresi\'ones cuant\'ilicas, que tienen la virtud de no requerir estacionariedad en el proceso generador de datos. Adem\'as de esto, la regresi\'on cuant\'ilica tiene la virtud de permitir modelar los cuantiles de la distribuci\'on de la variable explicada en t\'erminos de varias covariables, de forma que, en el contexto de este trabajo, se puede determinar la influencia de las covariables sobre la tasa de inflaci\'on \emph{en distintos extremos de su distribuci\'on}. En esta secci\'on primero se expone la teor\'ia estad\'istica b\'asica acerca de la regresi\'on cuant\'ilica y, luego, se explica la forma particular en la que esta se usa en este trabajo. 
\subsection{Regresi\'on cuant\'ilica}
Dados una v.a. $Y$ y vector $X=(X^{(1)},...,X^{(p)})$ de covariables los $\tau-$cuantiles incondicional y condicional de $Y$ dado $X=x \in \mathbb{R}^p$ se definen, respectivamente,
\begin{equation*}
\begin{split}
    Q(\tau)&=inf\{y: F_{Y}(y)\geq \tau\} \\
    Q(\tau \rvert X=x) &:= inf\{y: F_{Y\rvert X}(y\rvert x)\geq \tau\} 
\end{split}
\end{equation*}
Una propiedad fumdamental que cumplen los cuantiles es:
\begin{equation}\label{cuant_prop}
    Q(\tau)=\underset{c}{\text{argmin}}\mathbb{E}(p_\tau(Y-c))
\end{equation}
donde $\rho_\tau(u):=(\tau-\mathbbm{1}_{\{u<0\}})u$ es una funci\'on de distancia. Dada una muestra $((Y_i, X_i))$, no necesariamente independiente, el proceso generador de datos es
\begin{equation*}
    Q(\tau \rvert X_i)=\beta_0 \sum_{j=1}^p \beta_jX_i^j+\epsilon_i \forall i=1,...,n
\end{equation*}
para un vector de par\'ametros $\beta=(\beta_j)_{j=0}^p$ y t\'erminos de error $(\epsilon_i)_{i=1}^n$. El estimador utilizado en la literatura, as\'i como ac\'a, constituye una versi\'on muestral de (\ref{cuant_prop}):
\begin{equation*}
    \hat{\beta}= \underset{\beta}{\text{argmin}} \sum_{i=1}^n \rho_\tau(Y_i-\sum_{j=1}^p\beta_jX_i^{(j)})
\end{equation*}
Se puede realizar inferencia sobre los par\'ametros aprovechando que, incluso en presencia de errores dependientes y/o correlacionados, estos son asit\'oticamente normales y, por tanto, se pueden usar las pruebas usuales aplicadas en regresi\'on lineal. 
\subsection{Extremos de la tasa de inflaci\'on y regresi\'on cuant\'ilica}

Siguiendo a, \citeA{lopez_iar}, en este trabajo se estiman regresiones cuant\'ilicas para extremos de la distribuci\'on de la tasa de inflaci\'on. Estos autores estiman regresiones para los cuantiles $1\%, \ 50\% \ \text{y} \ 90\%$. Extendiendo su metodolog\'ia, ac\'a se estiman regresiones para los cuantiles $1\%, 2\%,...20\% $, a fin de estudiar la cola inferior de la distribuci\'on, y $80\%,...,99\%$, para la cola superior. Para la estimaci\'on se utiliza una base de datos con los valores de las covariables en su valor contempor\'aneo, un rezago en la tasa de inflaci\'on. Adem\'as, considerando lo discutido en la secci\'on \ref{descrip_extremos}, para la brecha del PIB se incluye el tercer y cuarto rezago y, para la inflaci\'on de materias, primas importadas, el primer y tercer rezago. 
Esto se realiza a fin de mitigar los efectos de alta multicolinealidad entre las covariables pero, a su vez, tomar en cuenta que muchas de las variables macroecon\'omicas como las ac\'a examinadas suelen tener efectos rezagados sobre la inflaci\'on, como se vio previamente en el an\'alisis descriptivo. Para cada cuantil se utiliza un criterio de selecci\'on de mejores subconjuntos en las que, de todas las posibles especificaciones del modelo dada la base, se elige aquella con un menor AIC. En cada regresi\'on se realizaron pruebas de significancia de los coeficientes usando los errores est\'andar estimados por el m\'etodo de Kernell de Powell que es robusto ante errores correlacionados y/o heterosced\'asticos \cite{koenker2005quantile}. 

\section{Resultados}
\begin{figure}[H]
\centering
\caption{\\[0.0001cm] \small \textbf{Coeficientes estimados por cuantil, cola inferior}}
    \begin{subfigure}[b]{0.3\textwidth}
        \centering
        \includegraphics[width=\textwidth]{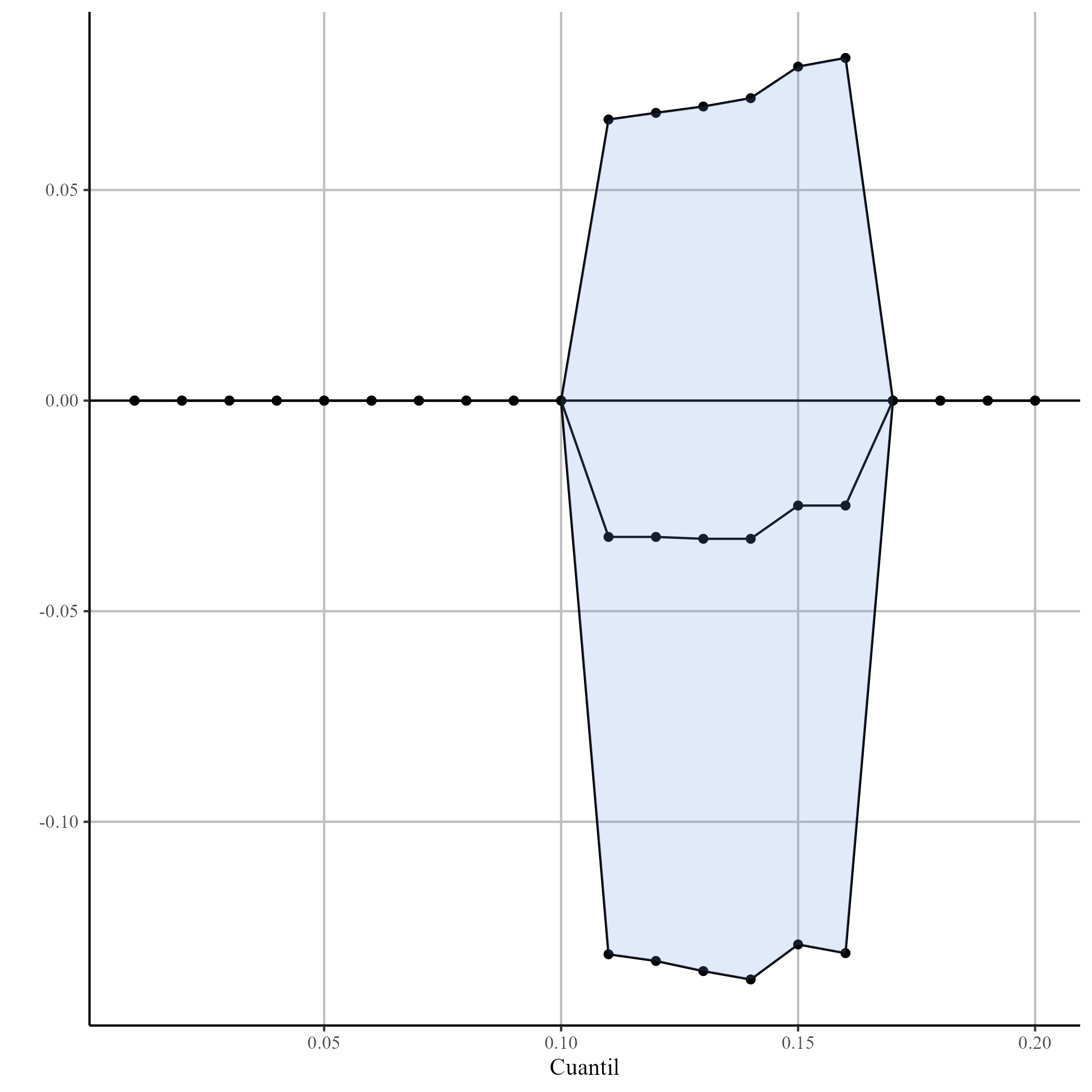}
        \caption{Brecha PIB}
        \label{fig:cf_infer_brecha_pib}                
    \end{subfigure}
    \hfill
    \begin{subfigure}[b] {0.3\textwidth}
    \centering
        \centering
        \includegraphics[width=\textwidth]{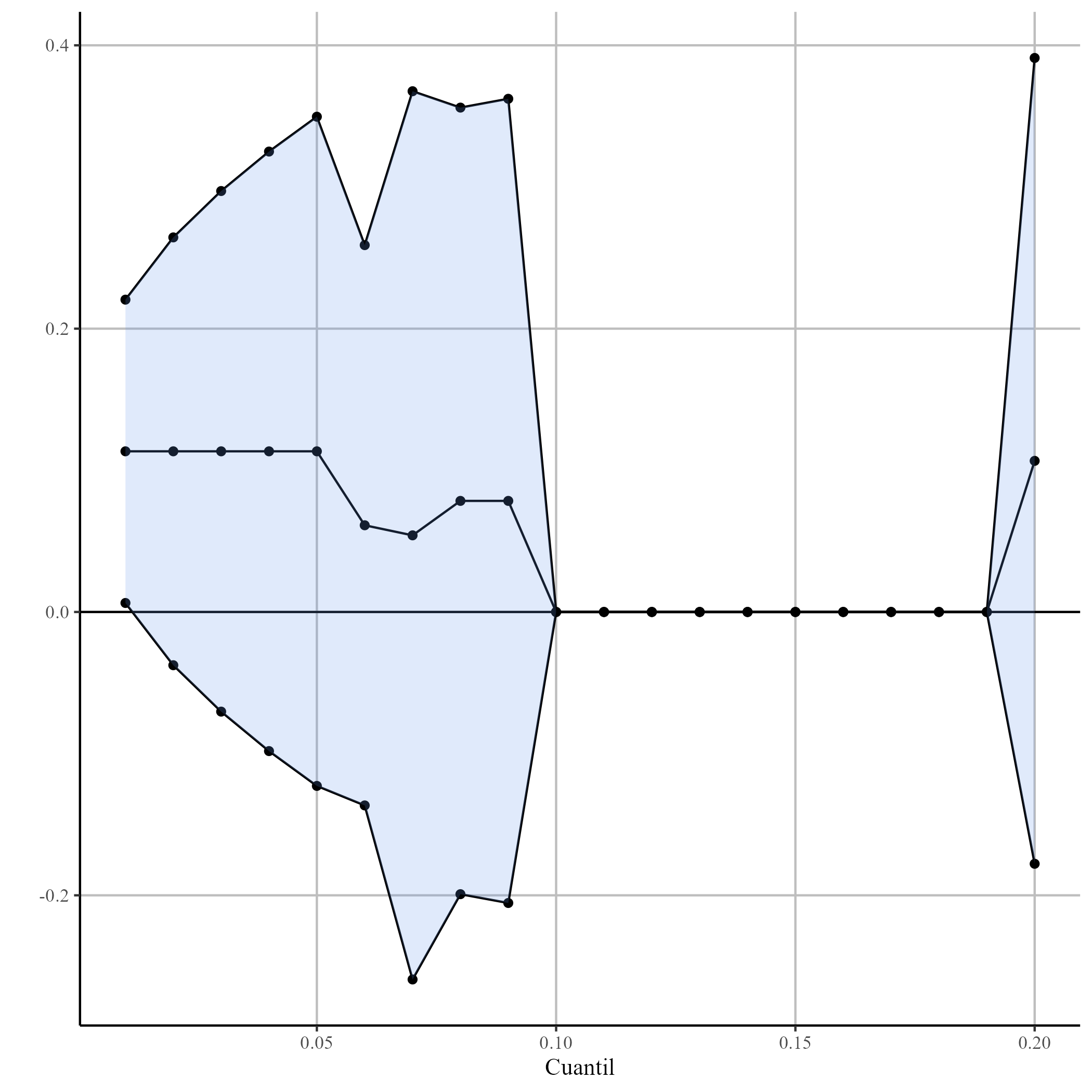}
        \caption{Expectativas}
        \label{fig:cf_infer_expect}
    \end{subfigure}
    \begin{subfigure}[b]{0.3\textwidth}
        \centering
        \includegraphics[width=\textwidth]{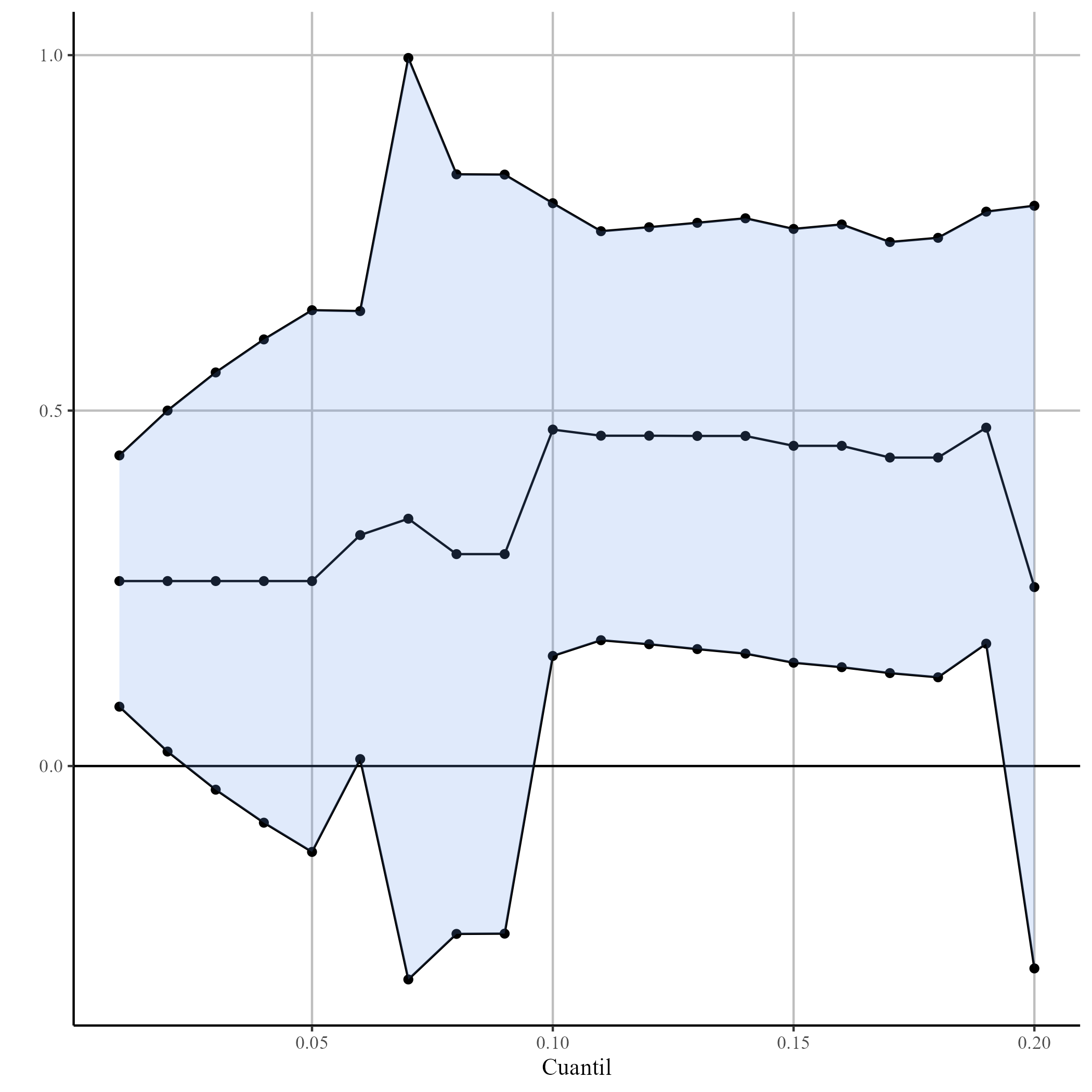}
        \caption{1º rezago, inflaci\'on}
        \label{fig:cf_infer_L1inf_ipc}          
    \end{subfigure}
    \\
    \begin{subfigure}[b]{0.3\textwidth}
        \centering
        \includegraphics[width=\textwidth]{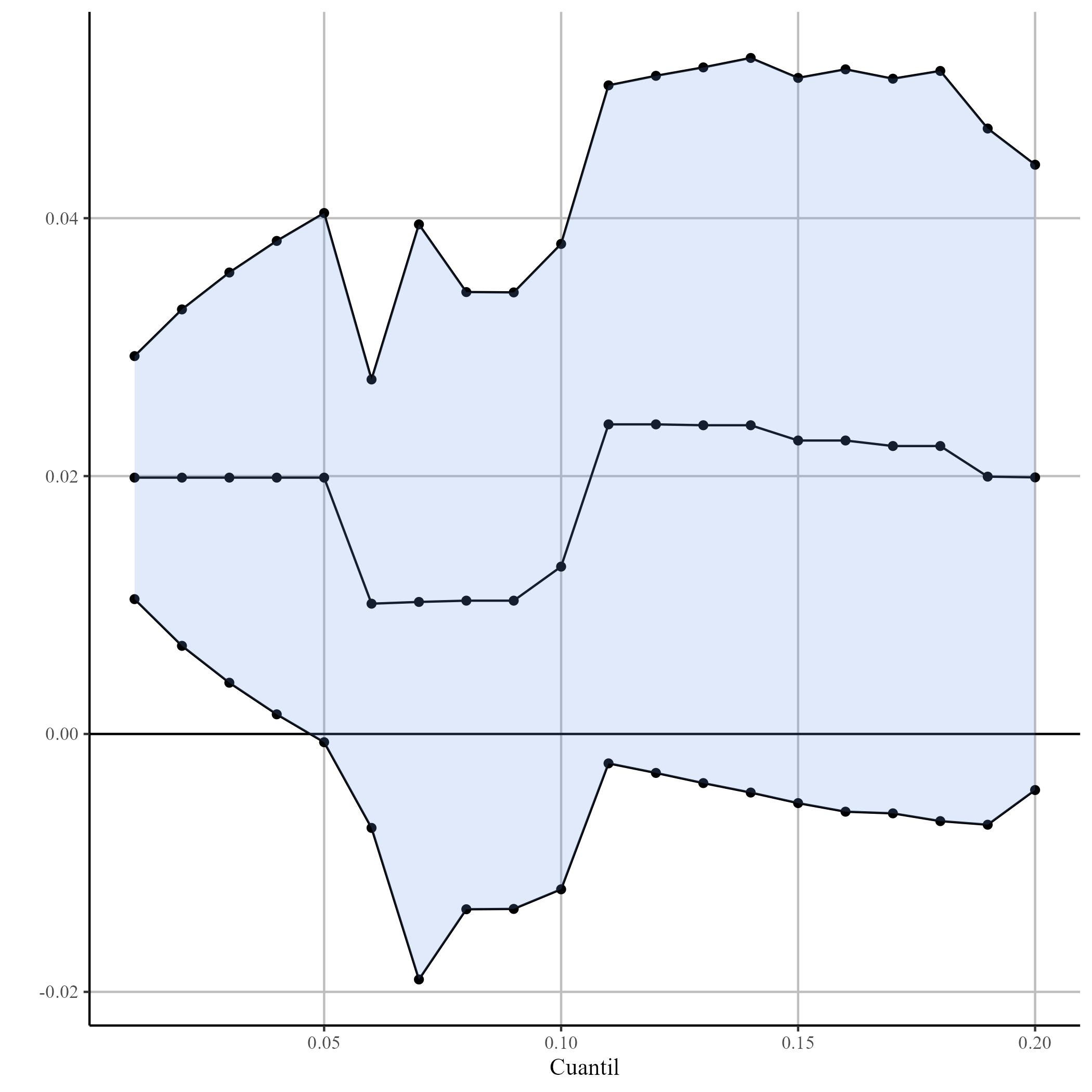}
        \caption{1º rezago, inflaci\'on importada}
        \label{fig:cf_infer_L1inf_import}                
    \end{subfigure}
    \hfill
    \begin{subfigure}[b] {0.3\textwidth}
    \centering
        \centering
        \includegraphics[width=\textwidth]{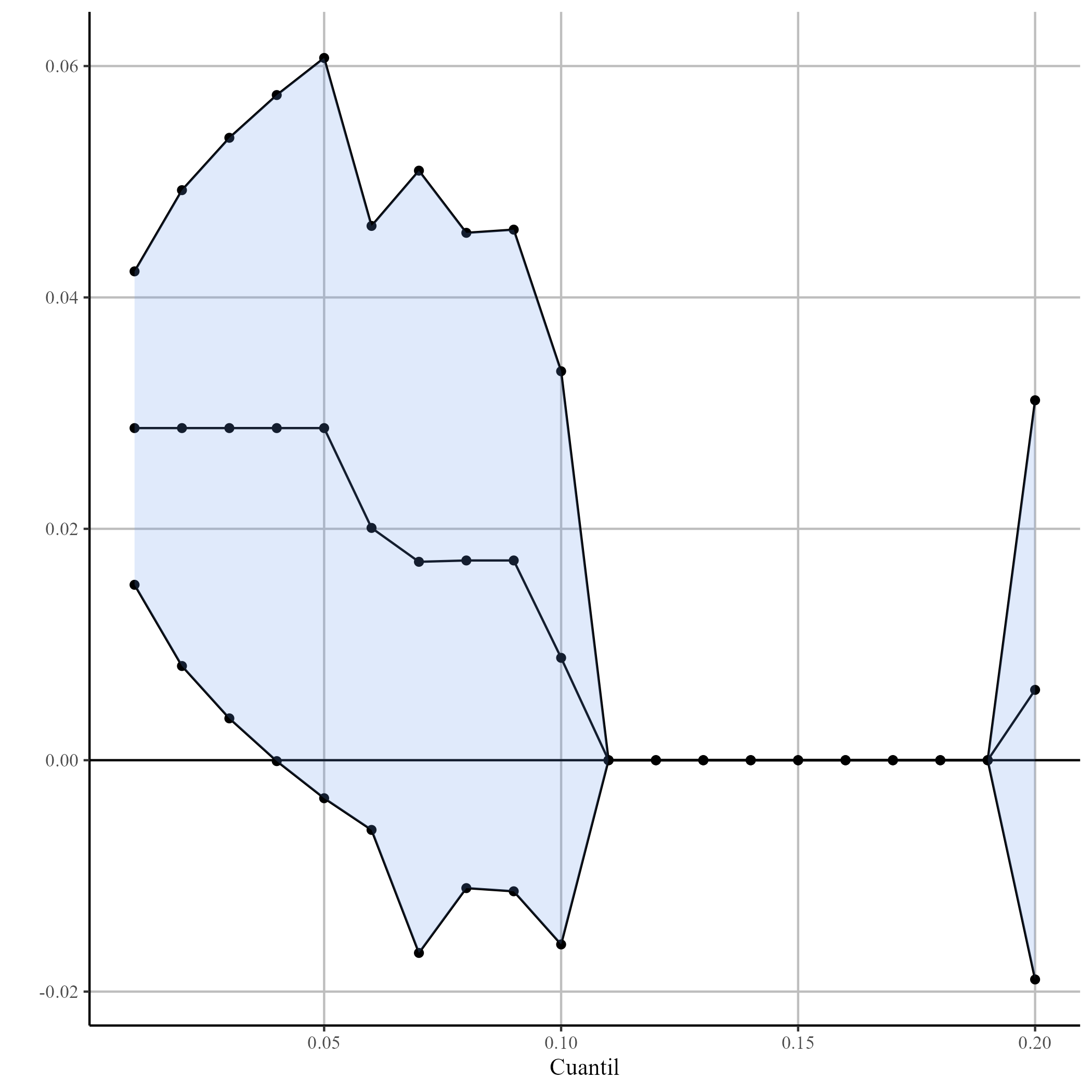}
        \caption{2º rezago, inflaci\'on importada}
        \label{fig:cf_infer_L2inf_import}
    \end{subfigure}
    \begin{subfigure}[b]{0.3\textwidth}
        \centering
        \includegraphics[width=\textwidth]{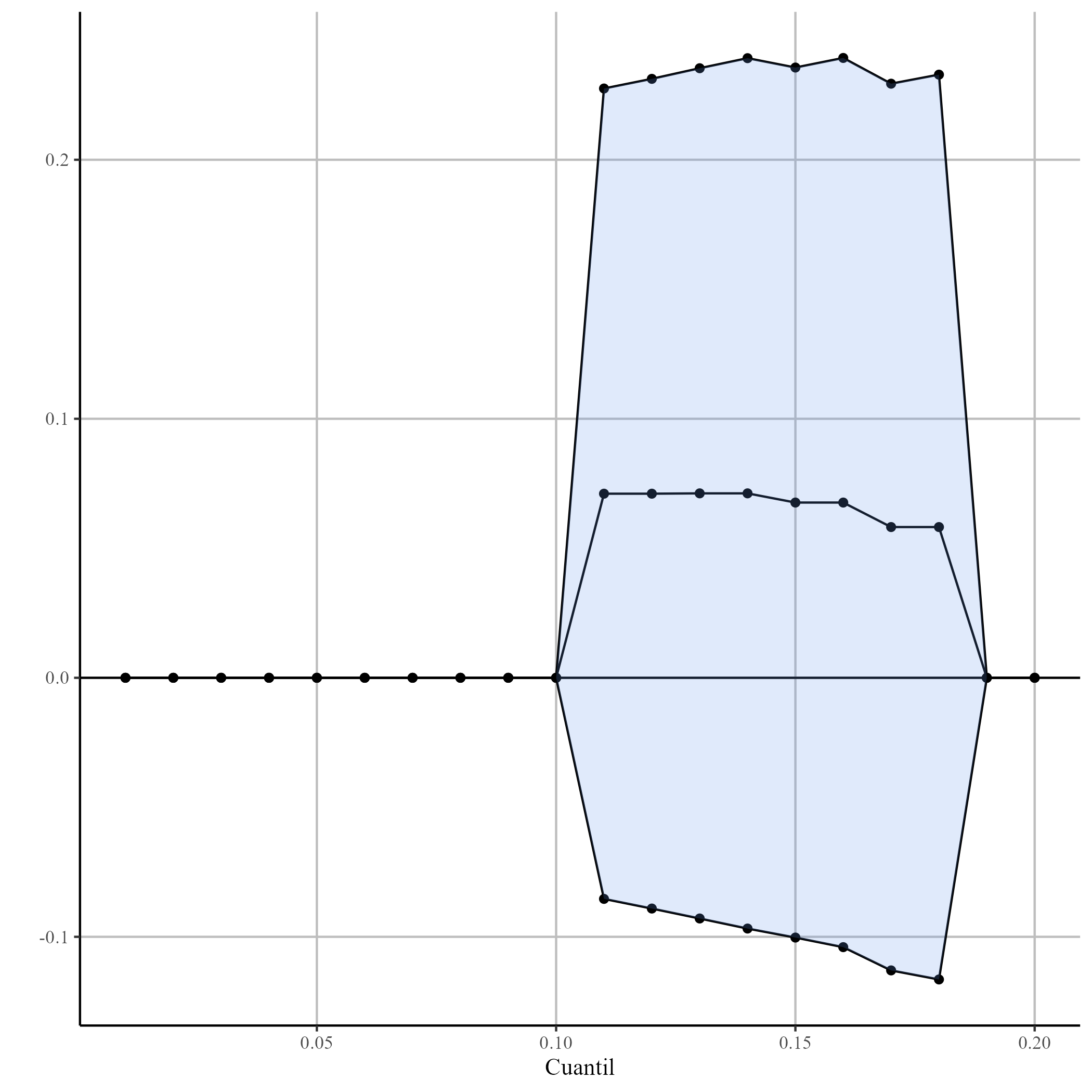}
        \caption{2º rezago, Brecha PIB}
        \label{fig:cf_infer_L3brecha_pib}                
    \end{subfigure}
    \label{cfs_inferior}
\end{figure}

 En la figura \ref{cfs_inferior} se muestran los coeficientes estimados e intervalos de confianza al 5\% para las covariables seleccionadas por cuantil de la cola inferior. Para los cuantiles del 1\% al 10\% las variables seleccionadas por el algorithmo desarrollado fueron las expectativas de inflación contemporaneas, primer y segundo rezago de la inflaci\'on importada, y primer rezago de la inflaci\'on; todas presentan coeficientes positivos, lo que sugiere que disminuciones en estas mueven la cola inferior de la distribuci\'on hacia la izquierda, haciendo m\'as probable eventos deflacionarios. Similarmente, en un\'isono con lo visto en la seccion \ref{descrip_extremos}, la inflaci\'on importada parece tener un efecto rezagado sobre los cuantiles inferiores, lo que puede explicar los periodos de deflaci\'on ocurridos cerca del 2015. N\'otese que, a pesar de que en general la brecha del PIB y su segundo rezago mantuvieron correlacion alta para los valores en general, en el caso de valores m\'inimos no aportaron valor, lo cual coincide con el hecho de que este gap fue cercano a cero en periodos de deflaci\'on y que la inflaci\'on parec\'ia no responder ante \emph{shocks} en la brecha del PIB (vea figura \ref{fig:brecha}), como se vio en el an\'alisis exploratorio, de manera que se entiende porqu\'e no desacelera los percentiles bajos de la inflación. Conforme nos alejamos de la cola inferior, i.e., del cuantil del 10\% en adelante, en su mayor\'ia, el segundo rezago de la inflaci\'on importada desaparece y la brecha del PIB entra al modelo, tanto en su valor contempor\'aneo y primer rezago. Los signos de estas dos coinciden con la posible interpretaci\'on, planteada en la secci\'on \ref{descrip_extremos}, de que \emph{shocks} a la baja en la brecha pueden tener efectos rezagados deflacionarios que la brecha contempor\'anea \say{intenta corregir}, si bien este efecto parece solo darse para valores no tan bajos en la tasa de inflaci\'on. 
 En general, la inflaci\'on e inflaci\'on importada rezagadas parecen ser importantes.
 
\begin{figure}[H]
\centering
\caption{\\[0.0001cm] \small \textbf{Coeficientes estimados por cuantil, cola superior}}
    \begin{subfigure}[b]{0.3\textwidth}
        \centering
        \includegraphics[width=\textwidth]{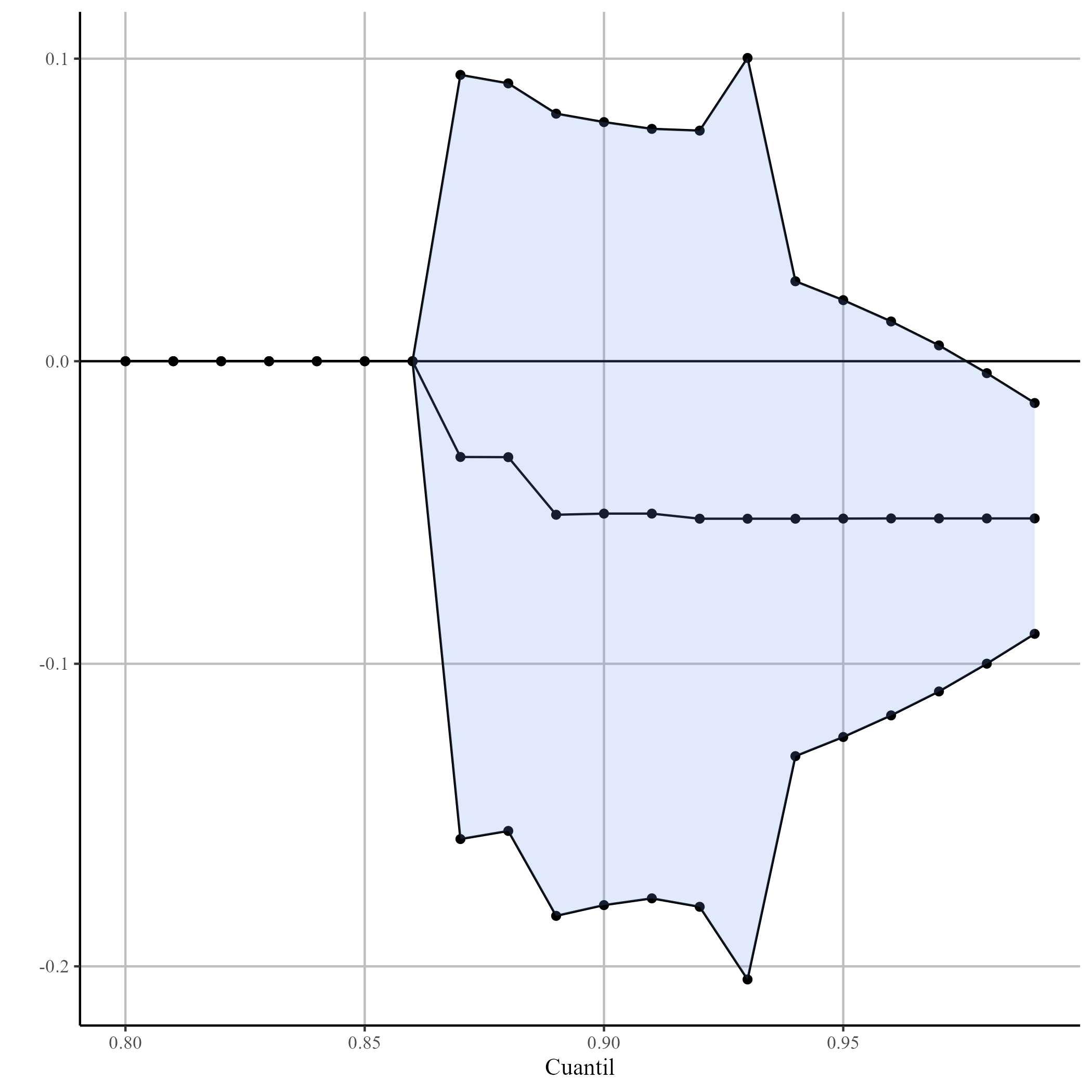}
        \caption{Brecha PIB}
        \label{fig:cf_super_brecha_pib}                
    \end{subfigure}
    \hfill
    \begin{subfigure}[b] {0.3\textwidth}
    \centering
        \centering
        \includegraphics[width=\textwidth]{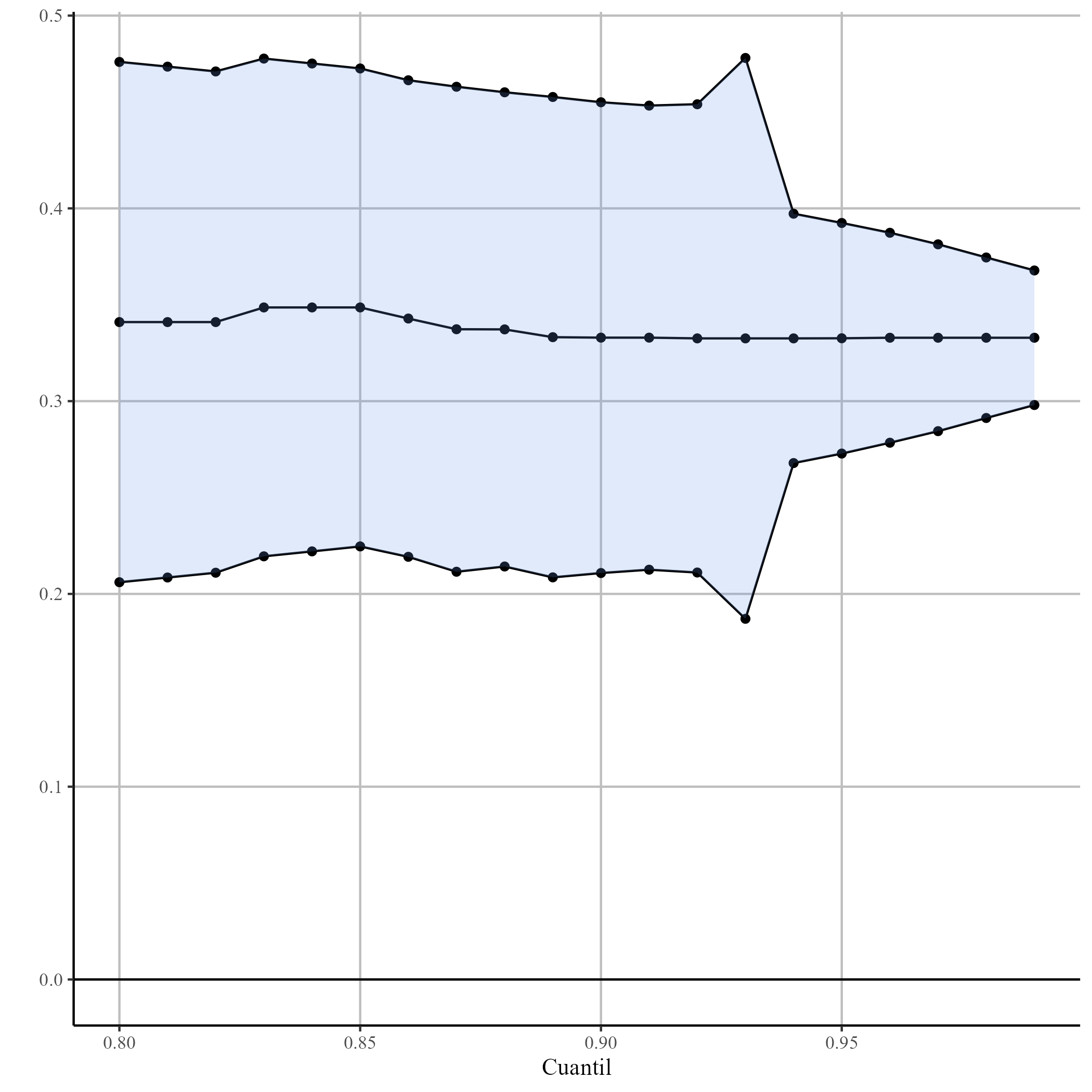}
        \caption{Expectativas}
        \label{fig:cf_super_expect}
    \end{subfigure}
    \begin{subfigure}[b]{0.3\textwidth}
        \centering
        \includegraphics[width=\textwidth]{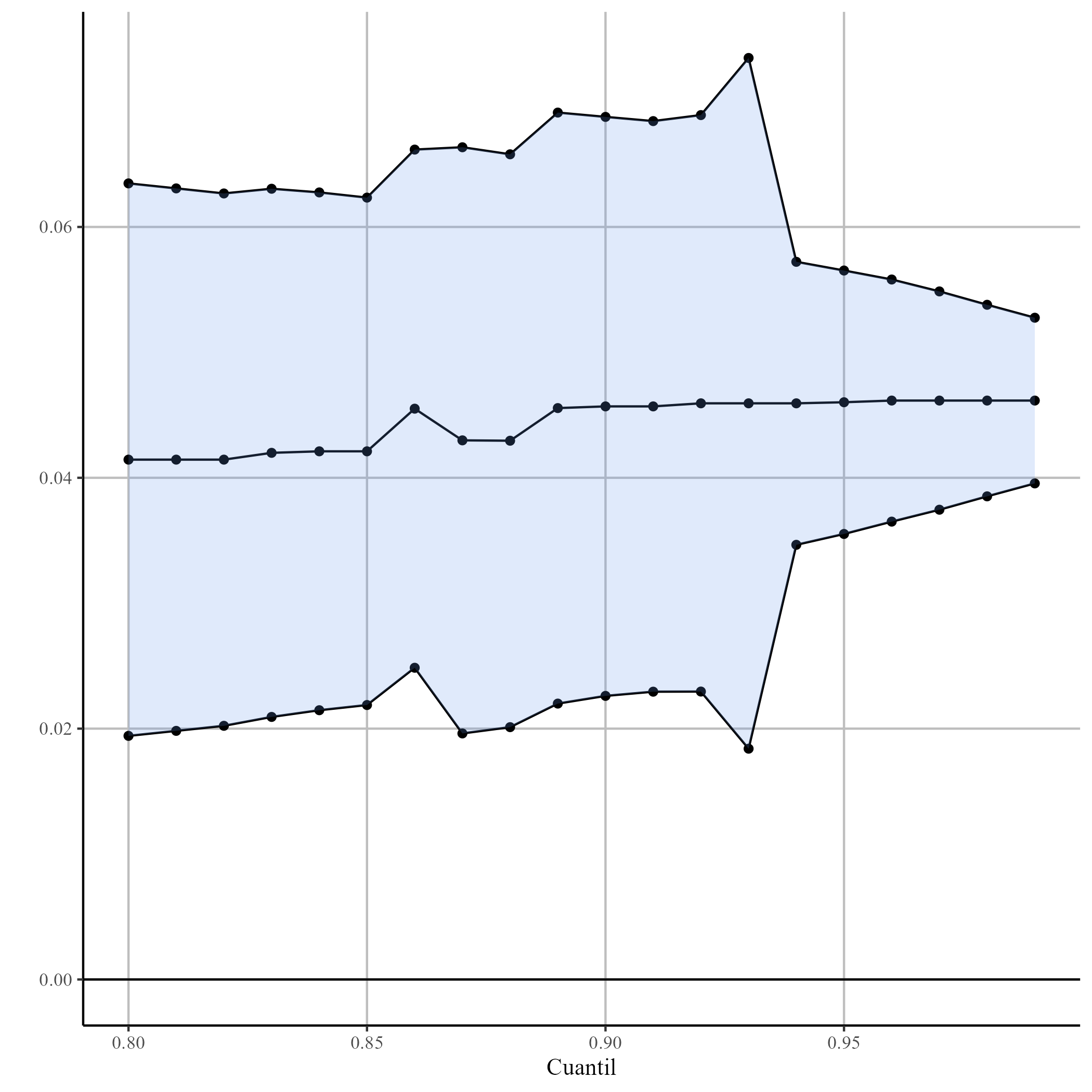}
        \caption{1º rezago, inflaci\'on importada}
        \label{fig:cf_super_L1inf_import}                
    \end{subfigure}
    \\
    \begin{subfigure}[b]{0.3\textwidth}
        \centering
        \includegraphics[width=\textwidth]{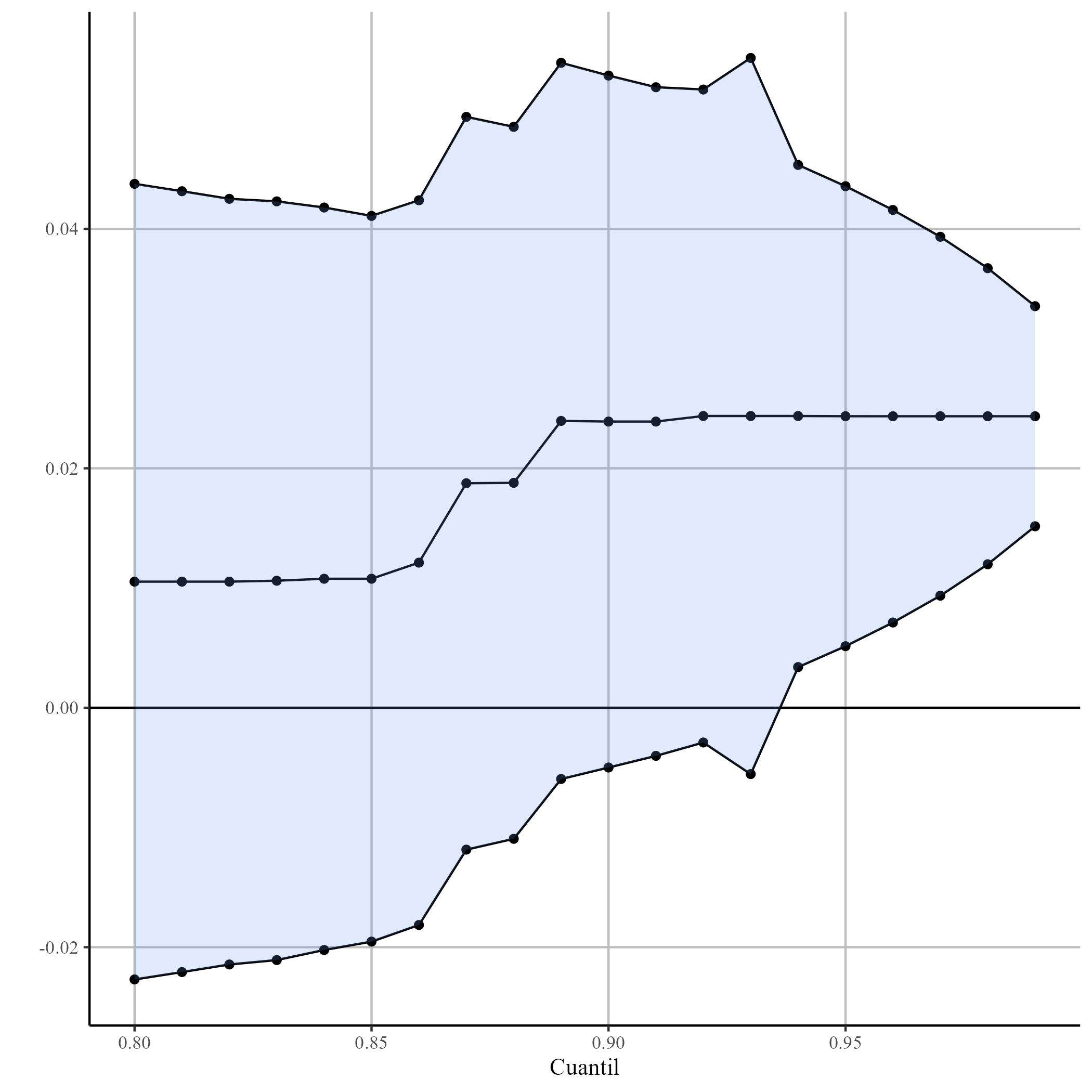}
        \caption{2º rezago, inflaci\'on importada}
        \label{fig:cf_super_L2inf_import}                
    \end{subfigure}
    \hfill
    \begin{subfigure}[b] {0.3\textwidth}
    \centering
        \centering
        \includegraphics[width=\textwidth]{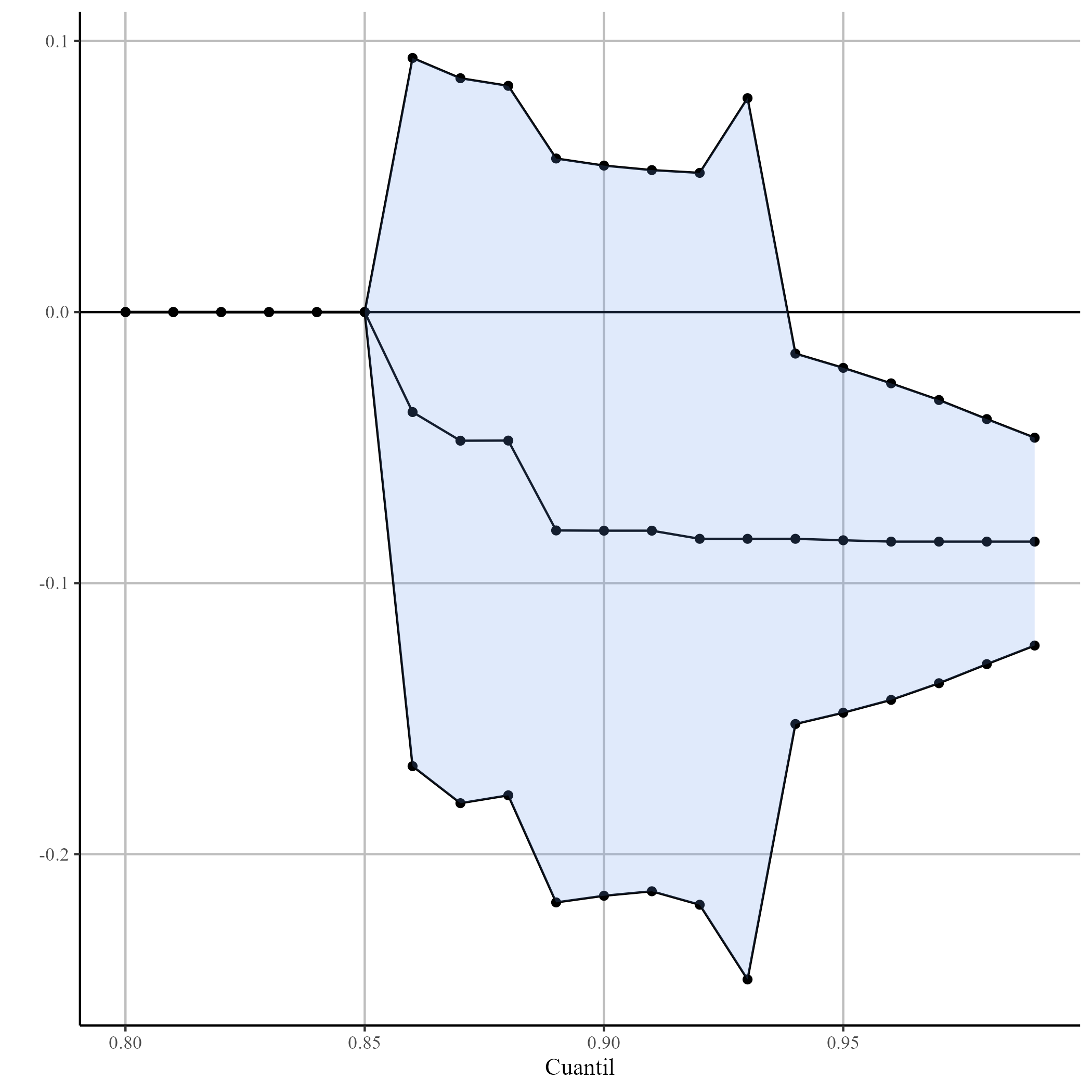}
        \caption{3º rezago, Brecha PIB}
        \label{fig:cf_super_L3brecha_pib}
    \end{subfigure}
    \begin{subfigure}[b]{0.3\textwidth}
        \centering
        \includegraphics[width=\textwidth]{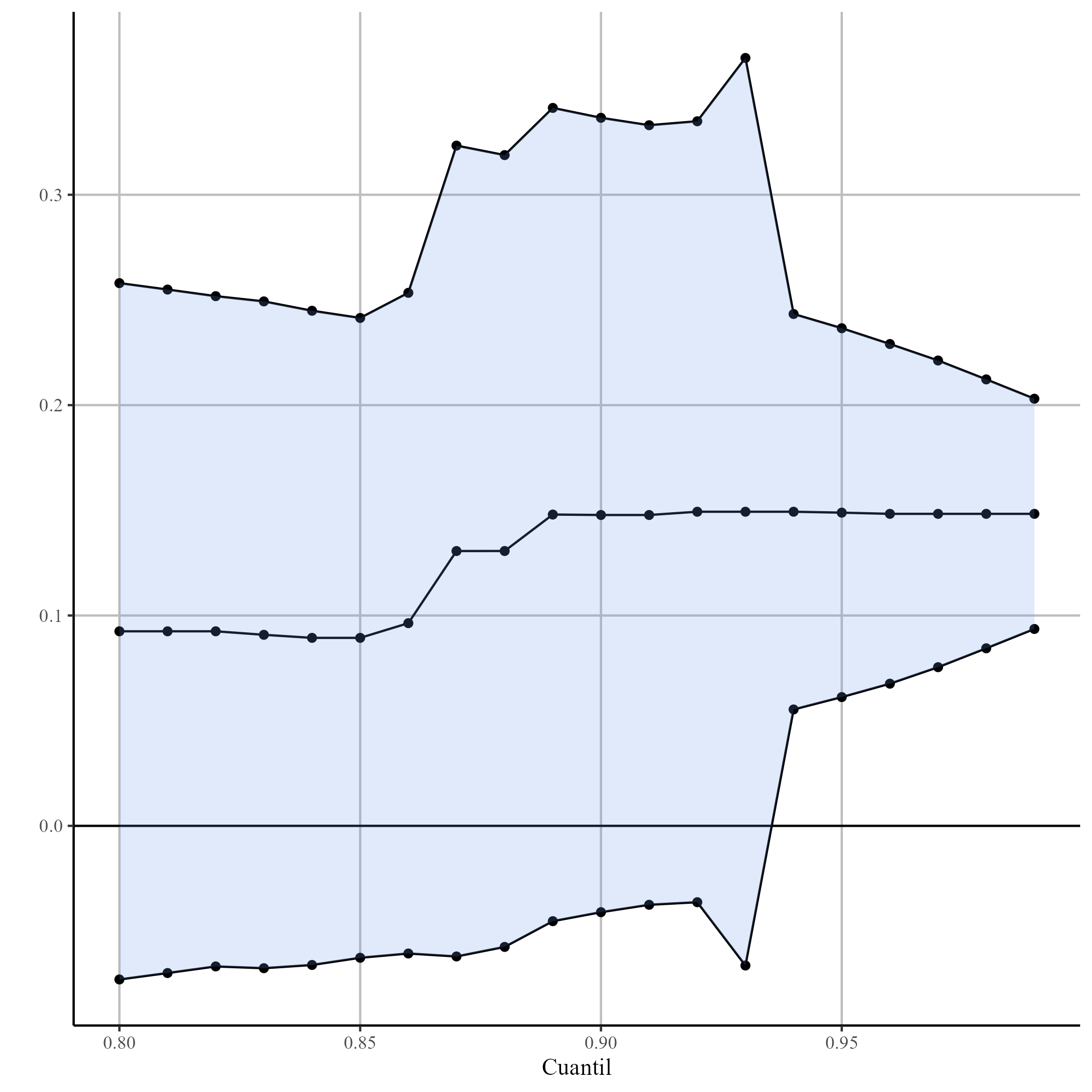}
        \caption{4º rezago, Brecha PIB}
        \label{fig:cf_super_L4brecha_pib}                
    \end{subfigure}
    \label{cfs_superior}
\end{figure}

Por otro lado, en la figura \ref{cfs_superior} se muestran los resultados para las colas superiores. Las expectativas, el cuarto rezago de la brecha del PIB, y los dos primeros rezagos de la inflaci\'on importada entran en todos los cuantiles y con coeficientes positivos, lo que coincide con las ideas te\'oricas. Sin embargo, el segundo y tercer rezago de la brecha entran negativos, para lo cual no encontramos una interpretaci\'on obvia. Es destacable que en ninguno de los cuantiles superiores entr\'o el rezago en la inflaci\'on, lo que sugiere que la probabilidad de valores altos en la tasa de inflaci\'on no est\'a relacionada con valores altos de la inflaci\'on previa, i.e., no hay inercia. Debe tomarse en cuenta que, si bien varias de las variables resultan no significativas, la prueba de significancia utilizada obedece a propiedades asint\'oticas y para un \'unico estimador del error est\'andar, por lo que deben tomarse con cautela, mientras que el AIC utilizado para la selecci\'on de modelo es m\'as confiable. 

\section{Conclusiones}
En esta investigaci\'on se busc\'o determinar la relaci\'on de varias variables macroecon\'omicas y los extremos de la tasa de inflaci\'on. Para ello se utilizaron regresiones cuant\'ilicas para varios cuantiles en las colas inferior y superior. Para los valores inferiores se obtuvo que la inflaci\'on de materias primas importadas tienen un efecto rezagado y positivo sobre la cola inferior de la distribuci\'on de manera que shocks a la baja de la primera mueve a la izquierda la cola inferior de la inflaci\'on deom\'estica en trimestres posteriores. Este fen\'omeno as\'i como las expectativas e infercia inflacionaria parecen ser las principales fuerzas que act\'uan sobre la cola inferior, manifest\'andose la brecha del PIB hasta cuantiles m\'as all\'a del 10\%. Para los valores superiores se obtuvo que las expectativas y rezagos en la inflaci\'on importada son las variables m\'as relevantes y tienen una relaci\'on positiva con los valores extremos de la inflaci\'on. Sin embargo, entra la brecha y uno de sus rezagos con coeficiente negativo, lo que dificulta la interpretaci\'on. Como limitaciones destacan tres caracter\'siticas del estudio. Primero, el filtro utilizado para computar el PIB potencial es el HP, por lo que los resultados podr\'ia no ser robustos ante otros m\'etodos. Segundo, la inferencia se realiz\'o apelando a propiedades asint\'oticas y bajo un \'unico estimador de los errores est\'andar. 

Dentro de las oportunidades de mejora se encuentra el utilizar el método de bootstrap que recomienda la literatura con el fin de obtener los intervalos de confianza sin requerir de supuestos adicionales

\bibliography{bibliografia}

\end{document}